\documentclass[twocolumn]{aastex631}

\usepackage{amsmath}
\usepackage{wasysym}
\usepackage{afterpage}
\usepackage{placeins}
\usepackage{multirow}

\shorttitle{Lunar MeV gamma-ray emission}
\shortauthors{Fujiwara et al.}

\begin{document}
\defcitealias{moskalenko_porter07}{MP07}
\title{The Moon as a Cosmic-Ray Spectrometer: Prospects for MeV Gamma-Ray Observations}

\correspondingauthor{Tatsuki Fujiwara}
\email{u000719k@ecs.osaka-u.ac.jp}

\author[0009-0002-9370-3313]{Tatsuki Fujiwara}
\affiliation{Department of Earth and Space Science, Graduate School of Science, The University of Osaka, 1-1, Machikaneyama, Toyonaka, Osaka 560-0043, Japan}

\author[0000-0003-1052-6439]{Ellis R. Owen}
\affiliation{Department of Earth and Space Science, Graduate School of Science, The University of Osaka, 1-1, Machikaneyama, Toyonaka, Osaka 560-0043, Japan}
\affiliation{Astrophysical Big Bang Laboratory (ABBL), RIKEN Pioneering Research Institute, Wak\={o}, Saitama, 351-0198 Japan}

\author[0000-0002-7272-1136]{Yoshiyuki Inoue}
\affiliation{Department of Earth and Space Science, Graduate School of Science, The University of Osaka, 1-1, Machikaneyama, Toyonaka, Osaka 560-0043, Japan}
\affiliation{Interdisciplinary Theoretical \& Mathematical Science Center (iTHEMS), RIKEN, 2-1 Hirosawa, 351-0198, Japan}
\affiliation{Kavli Institute for the Physics and Mathematics of the Universe (WPI), UTIAS, The University of Tokyo, 5-1-5 Kashiwanoha, Kashiwa, Chiba 277-8583, Japan}

\author[0000-0002-1853-863X]{Manel Errando}
\affiliation{Department of Physics, Washington University in St Louis, One Brookings Drive, St. Louis, MO 63130, USA}

\author[0000-0002-8883-9251]{Kohei Fukuda}
\affiliation{Department of Earth and Space Science, Graduate School of Science, The University of Osaka, 1-1, Machikaneyama, Toyonaka, Osaka 560-0043, Japan}

\author[0000-0003-2930-350X]{Kazuhiro Nakazawa}
\affiliation{Nagoya University, Furo-cho, Chikusa-ku, Nagoya, Aichi 464-8601, Japan}

\author[0000-0003-2670-6936]{Hirokazu Odaka}
\affiliation{Department of Earth and Space Science, Graduate School of Science, The University of Osaka, 1-1, Machikaneyama, Toyonaka, Osaka 560-0043, Japan}

\author[0009-0002-2796-6844]{Keigo Okuma}
\affiliation{Nagoya University, Furo-cho, Chikusa-ku, Nagoya, Aichi 464-8601, Japan}

\author[0000-0003-0120-1229]{Kentaro Terada}
\affiliation{Department of Earth and Space Science, Graduate School of Science, The University of Osaka, 1-1, Machikaneyama, Toyonaka, Osaka 560-0043, Japan}

\author[0000-0001-7209-9204]{Naomi Tsuji}
\affiliation{Faculty of Science, Kanagawa University, 3-27-1 Rokukakubashi, Kanagawa-ku, Yokohama-shi, Kanagawa 221-8686}
\affiliation{Interdisciplinary Theoretical \& Mathematical Science Center (iTHEMS), RIKEN, 2-1 Hirosawa, 351-0198, Japan}

\author[0009-0008-5304-5233]{Yasunobu Uchiyama}
\affiliation{Graduate School of Artificial Intelligence and Science, Rikkyo University, 3-34-1 Nishi Ikebukuro, Toshima-ku, Tokyo 171-8501, Japan}
\affiliation{Department of Physics, Rikkyo University, 3-34-1 Nishi Ikebukuro, Toshima-ku, Tokyo 171-8501, Japan}

\author[0000-0002-5345-5485]{Hiroki Yoneda}
\affiliation{The Hakubi Center for Advanced Research, Kyoto University, Yoshida Ushinomiyacho, Sakyo-ku, Kyoto 606-8501, Japan}
\affiliation{Department of Physics, Kyoto University, Kitashirakawa Oiwake-cho, Sakyo-ku, Kyoto 606-8502, Japan}
\affiliation{RIKEN Nishina Center, 2-1 Hirosawa, Wak\={o}, Saitama 351-0198, Japan}
\affiliation{Kavli Institute for the Physics and Mathematics of the Universe (WPI), UTIAS, The University of Tokyo, 5-1-5 Kashiwanoha, Kashiwa, Chiba 277-8583, Japan}

\author[0000-0001-8500-2416]{Ao Zhang}
\affiliation{Department of Physics, Washington University in St Louis, One Brookings Drive, St. Louis, MO 63130, USA}

\collaboration{13}{}

\author{}
\affiliation{}
\affiliation{}

\begin{abstract}
The Moon is the closest celestial gamma-ray emitting object. Its gamma-ray emission arises from interactions between Galactic cosmic rays (CRs) and the lunar surface. While the lunar GeV gamma-ray spectrum is dominated by a continuum from hadronic decay processes, the MeV emission exhibits both continuum and distinctive spectral lines from nuclear de-excitation and radioactive decay processes. Using {\tt Geant4} Monte Carlo particle simulations, we model the lunar gamma-ray spectrum. Our results demonstrate its consistency with \textit{Fermi}-LAT observations, and 
 predict that next-generation MeV gamma-ray instruments will detect both the lunar MeV continuum and several key spectral line features, notably the $1.779~\mathrm{MeV}$ line from $\mathrm{^{28}Si}$ de-excitation enhanced by the lunar surface composition, the $e^+e^-$ annihilation line, and radioactive decay lines from $\mathrm{^{22}Na}$ ($\tau\approx3.75\,\mathrm{yr}$) and long-lived $\mathrm{^{26}Al}$ ($\tau\approx1\,\mathrm{Myr}$). These gamma-ray lines are sensitive to CRs with energies $\lesssim1\,\mathrm{GeV\,nuc^{-1}}$, offering unique temporal probes of CR activity over different timescales. Observations of the lunar MeV gamma-ray spectrum will therefore open a new window to study the current irradiation of the solar-terrestrial environment by low-energy CRs and its long-term temporal evolution.
\end{abstract}


\section{Introduction} \label{sec:intro}

The Moon serves as a unique laboratory for the study of cosmic rays (CRs) through its gamma-ray emission. These gamma rays are produced by CR bombardment of the lunar surface layer \citep{morris84}, with the dominant component being a continuum from inelastic CR collisions. This gamma-ray emission provides a valuable probe of Galactic CR spectra around the Moon, allowing us to constrain the local interstellar spectrum (LIS) after accounting for the solar modulation effect.

The lunar gamma-ray continuum was first detected by the Energetic Gamma Ray Experiment Telescope (EGRET) on board the Compton Gamma Ray Observatory (CGRO) in the energy range of $50$--$500\,\mathrm{MeV}$  \citep{thompson+97}, and later studied in detail by the Large Area Telescope (LAT) on board the \textit{Fermi} Gamma-ray Space Telescope in the range of $30\,\mathrm{MeV}$--$2\,\mathrm{GeV}$ \citep{abdo+12, ackermann+16}. 
Through detailed Monte Carlo simulations, \citet{ackermann+16} demonstrated that the observed spectrum in the GeV band is well reproduced by models of Galactic CR interactions, establishing the Moon as an effective probe of the Galactic CR spectrum above hundreds of MeV energies per nucleon (hereafter nuc)\footnote{Gamma-ray emission from small solar system objects could also serve a similar role in the future \citep[see e.g.,][]{moskalenko+08, degaetano+23, siegert+24}.}.

Theoretical studies of lunar gamma-ray emission, particularly \citet{moskalenko_porter07} (hereafter MP07), predict that the spectrum should extend into the MeV regime and below. While GeV emission is dominated by continuum emission from neutral pion decay ($\pi^0\rightarrow\gamma\gamma$), the MeV band is expected to exhibit a rich spectrum of both continuum and discrete line features, including nuclear de-excitation lines, the $511\,\mathrm{keV}$ pair annihilation line, and signatures of various radioactive decays. Although MeV gamma-ray spectroscopy measurements have been attempted by lunar orbiters \citep{metzger+73, lawrence+98, hasebe+08, ma+08, ma+13}, conclusive observations of these features are yet to be achieved due to background challenges and instrumental effects, primarily arising from CR-induced nuclear interactions with the detectors themselves.

Lunar gamma-ray emission serves as a general tracer of Galactic CRs in the solar-terrestrial environment, particularly in the MeV energy range where direct CR observations are challenging. Moreover, MeV gamma-ray lines provide unique temporal information about CR activities. While the continuum emission, nuclear de-excitation lines, and the 511~keV annihilation line are produced almost instantaneously after a CR interaction and reflect the present CR intensity, radioactive decay lines originate from cosmogenic nuclides with varying half-lives. These decay lines therefore allow us to probe past  CR activities. By combining different radioactive decay lines, the history of local CR intensity variations over multiple timescales can be reconstructed.

Several next-generation MeV gamma-ray missions are planned for launch in the late 2020s to 2030s, including the Compton Spectrometer and Imager \citep[COSI;][]{tomsick+23}, the Gamma Ray and AntiMatter Survey \citep[GRAMS;][]{aramaki+20}, the All-sky Medium Energy Gamma-Ray Observatory eXplorer \citep[AMEGO-X;][]{fleischhack+22}, the Galactic Explorer with a Coded Aperture Mask Compton Telescope \citep[GECCO;][]{orlando+22}, enhanced ASTROGAM \citep[e-ASTROGAM;][]{deangelis+18, deangelis+21}, and GammaTPC \citep{shutt+25}. While MeV gamma-ray astronomy has historically been limited by significant instrumental backgrounds, these upcoming missions are designed to 
achieve sensitivities one to two orders of magnitude better than previous instruments such as the COMPTEL on board CGRO \citep{schonfelder+84} and the Spectrometer on INTEGRAL (SPI) \citep{vedrenne+03}. This significant improvement in sensitivity makes it timely to revisit predictions of the lunar MeV gamma-ray spectrum in preparation for these new observational capabilities.

In this work, we present new Monte Carlo simulations using the {\tt Geant4} toolkit \citep{agostinelli+03, allison+06, allison+16} to  model the lunar MeV-GeV gamma-ray spectrum induced by Galactic CRs, including various hadronic physics models. While Monte Carlo simulation is the only way to obtain robust predictions for such low-energy phenomena, variations persist in outcomes using different MeV hadron physics models. This is due to uncertainties in reaction cross-sections and differences in modeling frameworks. 

We arrange this paper as follows. 
In Section~\ref{sec:method}, we describe our calculation framework, including the adopted physics models and assumed CR energy spectra. 
In Section~\ref{sec:result}, we present our predicted gamma-ray spectra and demonstrate their consistency with the \textit{Fermi}-LAT observations in the GeV range. 
Section~\ref{sec:discussion} explores the prospects for detecting lunar MeV gamma-ray lines with next-generation instruments, focusing on how nuclear de-excitation lines and radioactive decay lines can probe both current and historical CR activities.
We summarize our findings in Section~ \ref{sec:conclusion}.

\section{Methods} \label{sec:method}

\subsection{Overview of Calculations} \label{subsec:method_overview}

We model the Moon as a sphere of radius $R_{\leftmoon} =1.737\times10^8\;\mathrm{cm}$ with a uniform-density surface layer, positioned at its semi-major axis between the Earth: $a_{\rm{EM}} = 3.844\times10^{10}\;\rm{cm}$. Details of the lunar surface model are provided in \S\ref{subsubsec:method_LunarModel}.

The gamma-ray emission from the Moon arises from CR-induced inelastic interactions with its surface layer. The lunar gamma-ray intensity $I_{\gamma}(E_{\gamma})$ can be expressed in terms of CR intensity  $I_{i}(E_i)$ for each species $i$ at the lunar surface:
\begin{align}
\label{eq:gamma_intensity}
    I_{\gamma}(E_{\gamma}) &= \sum_{i}\int\mathrm{d}E_i\,Y_{i}(E_{\gamma}|E_i)I_{i}(E_i),\\
    \label{eq:def_yield}
    Y_{i}(E_{\gamma}|E_i) &= \frac{1}{N_{i}(E_i)} \frac{\mathrm{d}N_{\gamma ,i}(E_{\gamma}|E_i)}{\mathrm{d}E_{\gamma}},
\end{align}
where $E_\gamma$ is the gamma-ray energy, $E_i$ is the \textit{i}-th CR kinetic energy per nucleon, and $Y_{i}(E_{\gamma}|E_i)$ is the differential gamma-ray yield. This yield is defined by the ratio of 
secondary gamma-rays emitted from the lunar surface $N_{\gamma, i}(E_{\gamma}|E_i)$ to the number of incident CRs $N_{i}(E_i)$. We assume isotropic CR intensity $I_{i}(E)$ in deriving Eq.(\ref{eq:gamma_intensity}), following \citet{ackermann+16}.
$I_{i}(E_{i})$ and $N_{i}(E_{i})$ are related by: 
\begin{equation}
\label{eq:relation_Ii&Ni}
    I_{i}(E_{i}) = \frac{1}{4\pi^2 R_{\leftmoon}^2 \Delta t} \frac{\mathrm{d} N_{i}(E_{i})}{\mathrm{d}E_{i}},
\end{equation}
where $\Delta t$ represents a discretized time interval over which CRs are injected.

Given the angular resolutions of MeV Compton telescopes \citep[e.g.,][]{tomsick+23, aramaki+20, fleischhack+22, orlando+22, deangelis+21, shutt+25}\footnote{The Doppler broadening effect limits the angular resolution of Compton telescopes $\lesssim1\,\mathrm{MeV}$. This lower limit of the resolution is $\gtrsim0.2\,\deg$ \citep[see][]{zoglauer_kanbach03}.}, the Moon can be treated as a point source. The observable lunar gamma-ray flux  is therefore: 
\begin{align}
\frac{\mathrm{d}N_{\gamma}(E_{\gamma})}{\mathrm{d}E_{\gamma}} &= \pi \left(\frac{R_{\leftmoon}}{a_{\mathrm{EM}}}\right)^2 I_{\gamma}(E_{\gamma})  \nonumber \\
\label{eq:gamma_flux}
    &= \pi \left(\frac{R_{\leftmoon}}{a_{\mathrm{EM}}}\right)^2 \sum_{i}\int\mathrm{d}E_i\,Y_{i}(E_{\gamma}|E)I_{i}(E_i).
\end{align}
We can predict the lunar gamma-ray spectrum by treating the gamma-ray yield $Y_{i}(E_{\gamma}|E_i)$ and CR intensity $I_{i}(E_i)$ as independent quantities in Eq.~\ref{eq:gamma_flux}. This formulation allows 
spectra for different CR environments around the Moon to be computed efficiently.

\subsection{Setup for Monte Carlo Simulations} \label{subsec:method_MCsetup}

To calculate the gamma-ray yields $Y_{i}(E_{\gamma}|E_i)$, which depend on the target composition, geometry, and nuclear interactions, we perform detailed Monte Carlo simulations using version 11.2.1 of the {\tt Geant4} toolkit\footnote{\url{https://geant4.web.cern.ch}} \citep{agostinelli+03, allison+06, allison+16}.

{\tt Geant4}  offers comprehensive particle transport simulation capabilities, including detailed modeling of geometry, material properties, particle tracking, and 
electromagnetic, hadronic, and optical physical processes. It has been used in particle physics studies \citep[e.g.,][]{aad+12, chatrchyan+12} and also in medical physics applications \citep[e.g.,][]{archambault+03, arce+21} and space research \citep[e.g.,][]{ackermann+12, odaka+18}.

\subsubsection{Lunar Surface Model} \label{subsubsec:method_LunarModel}

\begin{figure} [t]
    \plotone{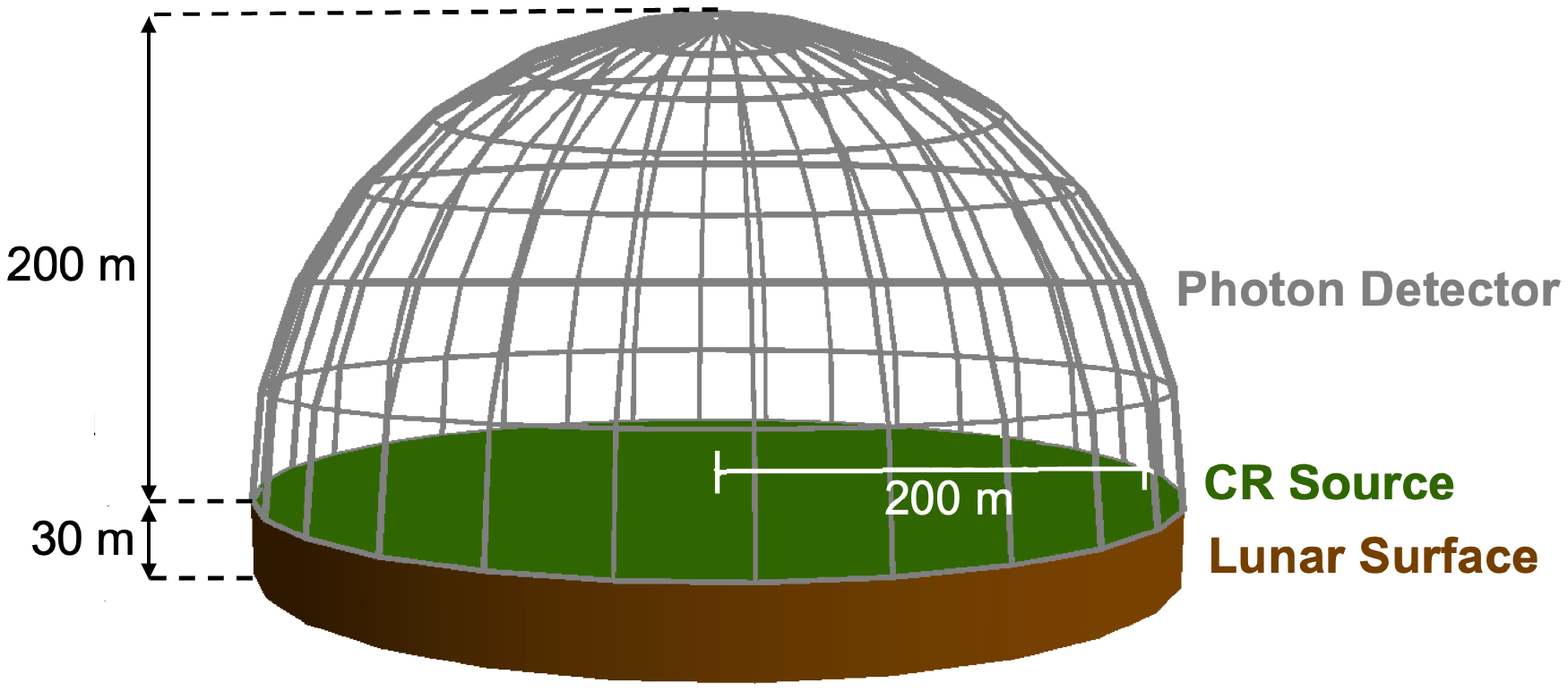}
    \caption{Schematic illustration of the model configuration, showing the lunar surface (brown; \S\ref{subsubsec:method_LunarModel}) covered by a source of primary CRs positioned directly on top of it (green; \S\ref{subsubsec:method_SimParam_CR}) to emulate irradiation by the Galactic CR flux. The CRs interact within the lunar surface region, down to a depth of 30 m, producing secondary gamma-rays. Some of these gamma-rays escape the surface. To measure this emission, a hemispherical shell with an inner radius of 200 m and a thickness of 1 m is defined (gray), within which the escaping secondary gamma-rays are counted.}
    \label{fig:g4geometry}
\end{figure}

Following \citetalias{moskalenko_porter07}, we model the lunar surface with a uniform density of $1.80\,\mathrm{g\,cm^{-3}}$ composed of 45\% $\mathrm{SiO_{2}}$, 22\% $\mathrm{FeO}$, 11\% $\mathrm{CaO}$, 10\% $\mathrm{Al_{2}O_{3}}$, 9\% $\mathrm{MgO}$, and 3\% $\mathrm{TiO_{2}}$ by weight, which corresponds to the elemental composition of lunar mare basalt \citep[e.g.,][]{lawrence+98, anand+03, prettyman+06}.
The nuclear interaction length in this 
medium, defined as the mean distance traveled by a hadronic CR before it undergoes an inelastic nuclear interaction, is $\approx57\,\mathrm{cm}$. The surrounding space is modeled with the GEANT4 ``G4\_GALACTIC'' material model with a density of $1\times10^{-25}\,\mathrm{g\,cm^{-3}}$ consisting of hydrogen only. 

Instead of the chemical composition above, we also examined an alternative mixture: 45\% $\mathrm{SiO_{2}}$, 6\% $\mathrm{FeO}$, 15\% $\mathrm{CaO}$, 26\% $\mathrm{Al_{2}O_{3}}$, and 8\% $\mathrm{MgO}$ by weight, which corresponds to the highland soils and the overall average composition of the highlands\footnote{The highlands form the remaining part of the lunar maria; in the highland model, the nuclear interaction length is $\approx55\,{\rm cm}$.} \citep[e.g.,][]{prinz+73, reid74, taylor75}. However, the outcome remained largely the same. We therefore adopt the 
lunar mare basalt model for the calculations in this work.

Following the set-up described by \citetalias{moskalenko_porter07}, we simulate a representative section of the lunar surface using a cylindrical target of radius $200\,\mathrm{m}$ and height $30\,\mathrm{m}$. Primary CRs, interacting in the lunar surface to a maximum depth of 30 m, generate secondary gamma-rays, some of which escape from the surface. To measure this gamma-ray emission, we define a hemispherical shell above the lunar surface, where secondary gamma-rays emerging from the surface are counted (Figure \ref{fig:g4geometry}). The dimensions of both the cylindrical lunar surface region and the hemispherical detection volume are chosen to be much larger than the nuclear interaction length, ensuring that particle cascade development is accurately captured in our simulation.

\subsubsection{{\tt Geant4} Physical Models}\label{subsubsec:method_HadronModel}

{\tt Geant4} provides a selection of Physics Lists. These represent alternative choices of particle-matter interaction models at different energies, with each list representing a particular set of models that are optimized for specific physical processes. The available physical models track the processes relevant to a wide range of particles: photons, (anti-)leptons, (anti-)baryons, (anti-)mesons, ions, anti-deuteron, anti-triton, anti-$\mathrm{^{3}He}$, and anti-alpha particles (anti-$\mathrm{^{4}He}$). They share identical implementations of electromagnetic processes (ionization, the photoelectric effect, Compton scattering, Coulomb scattering, multiple scattering, bremsstrahlung, and positron annihilation), decay processes (e.g., $\pi^{0}\rightarrow \gamma\gamma$), and radioactive decay processes (alpha and beta decay, electron capture). 

The key differences between Physics Lists lie in their treatment of hadronic interactions, particularly at low energies ($0 \leq E\leq 6\,\mathrm{GeV\,nuc^{-1}}$). Here, we provide a brief description of the hadronic interaction models in these Physics Lists that are relevant to protons, neutrons, and alpha particles.

At higher energies, all Physics Lists converge to use common models: the Fritiof parton (FTF) model \citep[$3 \leq E \leq 25\,\mathrm{GeV\,nuc^{-1}}$;][]{andersson+87, nilsson-almqvist_stenlund87} and the Quark-gluon String (QGS) model \citep[$E \ge 12\,\mathrm{GeV\,nuc^{-1}}$;][]{kaidalov82, kaidalov_ter-martirosyan82, capella+94}. The Precompound model \citep[${\rm 100\,keV\,nuc^{-1}} \leq E \leq {\rm 30\,MeV\,nuc^{-1}}$;][]{gudima+83} handles nuclear de-excitation in combination with these high-energy models, though the Bertini cascade model employs simpler precompound and de-excitation schemes.

The choice of Physics List can significantly impact the predicted gamma-ray yields, particularly in the MeV range where nuclear processes dominate. In this work, we adopt the Binary Cascade \citep[BIC;][]{folger+04} approach as a baseline model family. This offers a balance between computational efficiency and numerical accuracy for low-energy nuclear interactions. The standard \texttt{QGSP\_BIC} model provides reliable handling of nuclear processes relevant to radiation physics and particle transport. 
Built on \texttt{QGSP\_BIC}, \texttt{QGSP\_BIC\_HP} incorporates the high-precision (HP) neutron model below 20 MeV for enhanced treatment of neutron inelastic scattering, capture, and thermal neutron processes. These treatments are driven by the NeutronHP model using the G4NDL4.7 database \citep{mendoza+14}\footnote{See also \url{https://www-nds.iaea.org/geant4/} and references therein.}.
Further refinement is provided by \texttt{QGSP\_BIC\_AllHP}, which extends the HP approach beyond neutrons to also include protons, deuterons, tritons, $^{3}$He, and alpha particles below 200 MeV, though at increased computational cost. In this treatment, neutron interactions are handled by the NeutronHP model, while all other particles are treated using the ParticleHP model, which draws from the G4TENDL1.4 database, primarily based on TENDL-2019 with additions from ENDF/B-III.0 and JENDL/DEV-2020 \citep{koning+19, koning_rochman12, brown+18, nakayama+21}. It provides detailed handling of low-energy processes including spallation, nuclear excitation, evaporation, and fission. This approach is particularly valuable for studying nuclear de-excitation and radioactive decay processes relevant to lunar gamma-ray production. 

Alpha particle interactions are handled by the Binary Light Ion Cascade (BIC) model \citep[$0\leq E\leq 6\,\mathrm{GeV\,nuc^{-1}}$;][]{folger+04}, and the FTF model ($3\,\mathrm{GeV\,nuc^{-1}} \leq E \leq 100\,\mathrm{TeV\,nuc^{-1}}$) in all Physics Lists except \texttt{QGSP\_BIC\_AllHP}, which uses the ParticleHP model below 200 MeV.

For comprehensive model comparison, we also examine alternative approaches through the Bertini cascade \citep{bertini63, guthrie+68, wright_kelsey15} and INCL++ implementations \citep{boudard+13, davide+14, bennaceur_dobaczewski05, rodriguez+17}. The \texttt{QGSP\_BERT} and \texttt{QGSP\_BERT\_HP} variants employ the Bertini cascade model, which is particularly efficient for high-energy physics studies. Instead of individually tracking interactions between primary or secondary particles and target nucleons, as the BIC does, the Bertini cascade treats target nucleons as a gas with a Fermi momentum distribution, thereby providing its simpler de-excitation and precompound scheme. The \texttt{QGSP\_INCLXX} and \texttt{QGSP\_INCLXX\_HP} variants use the INCL++ model, offering detailed treatment of nuclear spallation and transmutation processes. As with the BIC variants, the HP suffix indicates enhanced neutron transport capabilities through the NeutronHP model.

Further details on these Physics Lists can be found in the {\tt Geant4} Physics Reference Manual and Physics List Guide. 
We compare the results of other model choices through detailed analysis of gamma-ray spectra in \S\ref{subsec:disc_InteractionModel}.

\subsubsection{Simulation Parameters for Physics Models} \label{subsubsec:method_SimParam_PhysModel}

For all our calculations, we set a maximum time threshold of $1.0\times10^{60}\,\mathrm{yr}$ for radioactive decay. This is sufficient to capture all radioactive decay processes relevant to our calculations\footnote{See Section 5.2.5 in `Book For Application Developers'' provided in the {\tt Geant4} Documentation.}. 
To avoid infrared divergence, where excessive photon production occurs as photon energy approaches zero (as e.g. in the bremsstrahlung process), we apply range cuts of $0.7\,\mathrm{mm}$ for secondary $e^{\pm}$ and photons in our calculations, following the default settings of the adopted Physics Lists. 
This corresponds to energy thresholds in lunar surface material of $\sim354\,\mathrm{keV}$ and $\sim5.2\,\mathrm{keV}$, respectively. Below these range cuts, secondary particles deposit their energy locally rather than being transported, thereby ensuring energy conservation. These choices do not significantly affect our results.
Hadronic processes use $0\,\mathrm{mm}$ proton range cuts to properly handle elastic processes and subsequent radioactive decay of recoil nuclei.

Unlike other transport models, the HP model violates energy and momentum conservation for certain processes but is instead designed to preserve overall average quantities (e.g., energy release, number of secondary particles).
In our HP simulations, cross sections of isotopes not present in the database are set to zero, and we set the parameter ``DO\_NOT\_ADJUST\_FINAL\_STATE'' to \textit{true} in order to prevent artificial gamma-ray generation, which results from the model attempting to satisfy energy and momentum conservation.

\subsubsection{Simulation Parameters for Primary CRs} \label{subsubsec:method_SimParam_CR}

Primary CR injections are limited to a solid angle of $2\pi\,\mathrm{sr}$ due to the target geometry. We enforce this in our simulations by using a circularly distributed CR source of radius $200\,\mathrm{m}$ centered on the lunar surface (Figure \ref{fig:g4geometry}), which generates particles following a cosine law to maintain CR isotropy.

We calculate gamma-ray yields for mono-energetic primary CRs with kinetic energies logarithmically binned from $1\,\mathrm{MeV\,nuc^{-1}}$ to $\approx200\,\mathrm{GeV\,nuc^{-1}}$. The upper energy limit is sufficient because the CR flux decreases steeply with energy $\propto E_i^{-2.7}$, making higher-energy contributions negligible. This was verified using the {\tt QGSP\_BERT} model.

\subsection{Galactic CRs and Solar Modulation} \label{subsec:method_CR}

\begin{figure} [t]
    \plotone{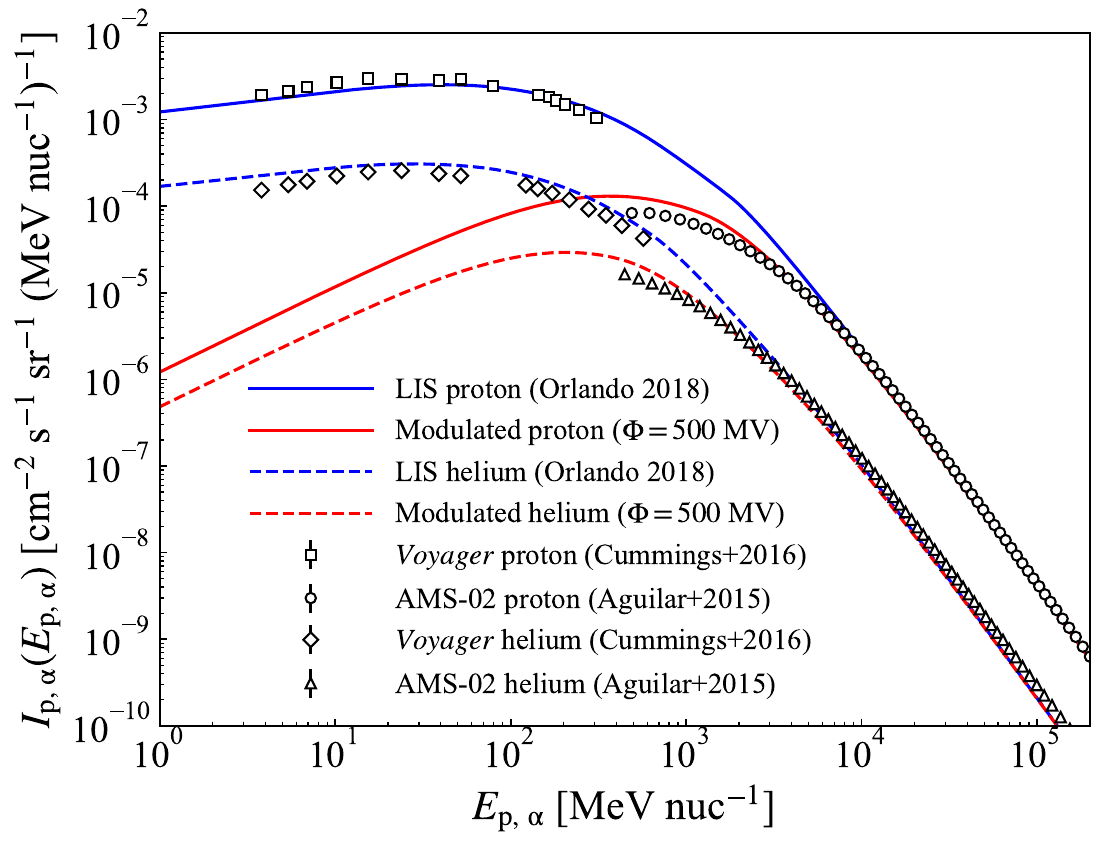}
    \caption{Galactic CR spectra for protons (solid line) and alpha particles (dashed line) are shown for scenarios with unmodulated LIS (blue) and medium modulation (red; $\Phi=500\,\mathrm{MV}$). LIS are constructed from \citet{orlando18}. Data points show measurements from {\it Voyager 1} \citep[squares and diamonds;][]{cummings+16} and AMS-02 \citep[circules and triangles;][]{aguilar+15_p, aguilar+15_a}. AMS-02 data above are taken from 2011 May 19 to 2013 November 26.}
    \label{fig:CRspec}
\end{figure}

\begin{figure*}[t!]
    \centering
    \includegraphics[width=0.45\linewidth]{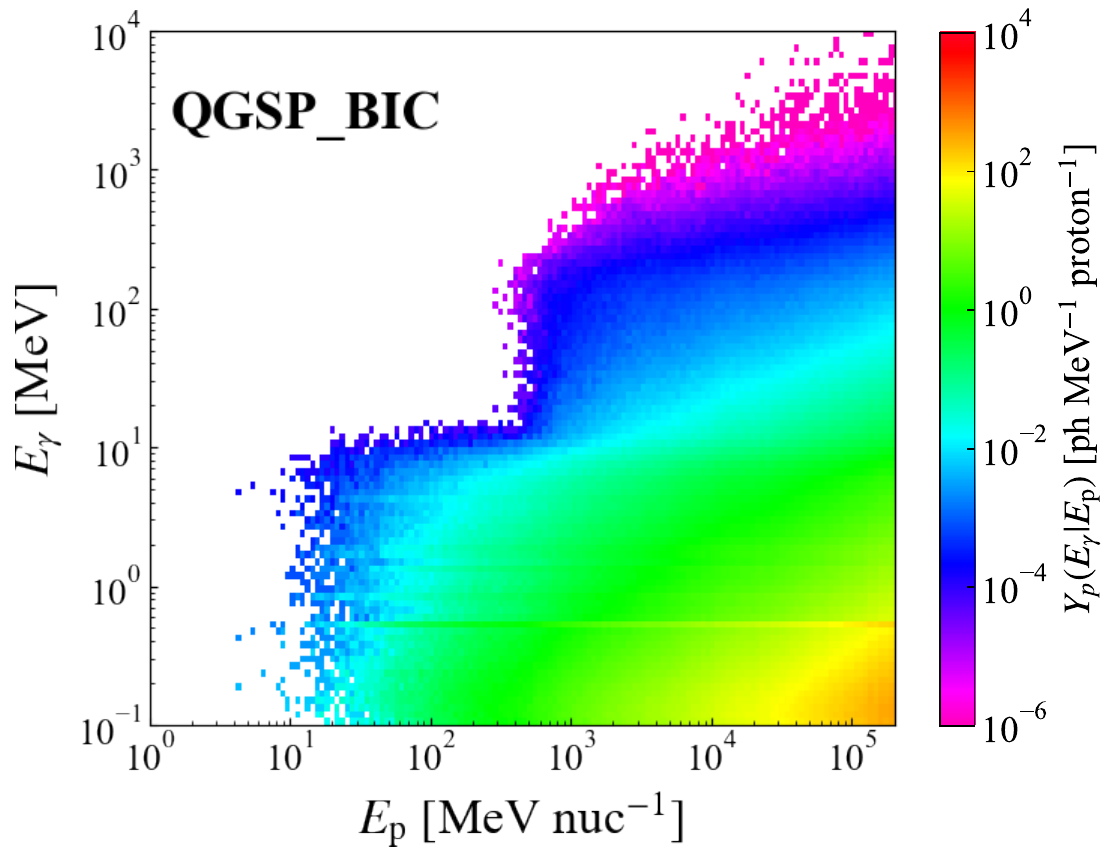}
    \hspace{0.05\linewidth}
    \includegraphics[width=0.45\linewidth]{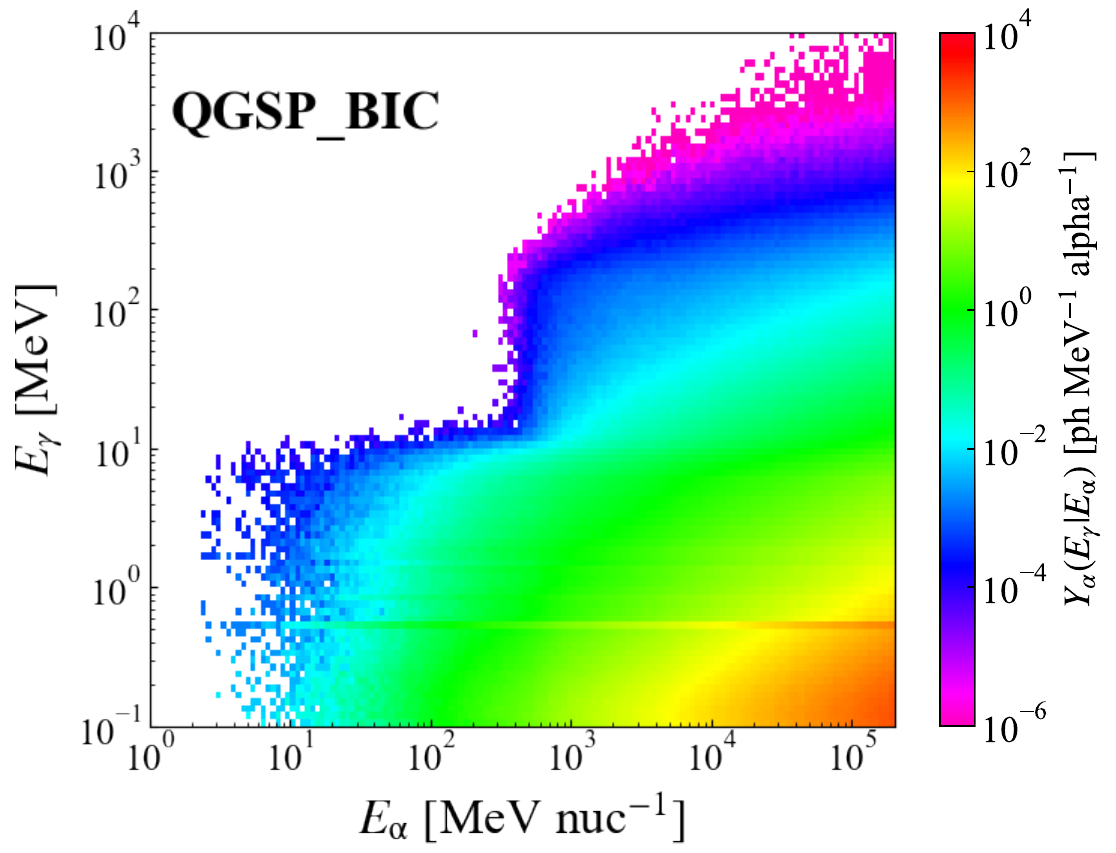}
       \caption{Color maps of the gamma-ray yields from Geant4 simulations for the {\tt QGSP\_BIC} model. The left and right panels show the results for proton and alpha injection, respectively. In both panels, the 511~keV line from electron-positron annihilation appears as a distinct horizontal feature. The gamma-ray energy $E_{\gamma}$ is divided into 100 equally spaced logarithmic scale bins from $0.1\,\mathrm{MeV}$ to $10\,\mathrm{MeV}$. }
   \label{fig:Gamyield}
\end{figure*}

To calculate the lunar gamma-ray spectrum $\mathrm{d}N_{\gamma}/\mathrm{d}E_{\gamma}$, we specify the CR intensity $I_{i}(E)$ around the Moon. Galactic CRs at Earth 
are dominated by protons and helium nuclei {(e.g., \citealt{lund84})}. For simplicity, we therefore consider only protons ($\mathrm{p}$) and alpha particles ($\mathrm{\alpha}$) as primary CR species in our calculations. We also account for the effect of solar modulation using the force field approximation \citep{gleeson_axford68}:
\begin{align}
        I_{i}(E_i) &= I_{i}^{\mathrm{LIS}}(E_i+e\Phi Z_{i}/A_{i}) \nonumber \\
        &\times\frac{(E_i+m_{i}c^2)^2 -(m_{i}c^2)^2}{(E_i+m_{i}c^2+e\Phi Z_{i}/A_{i})^2 -(m_{i}c^2)^2}, 
    \label{eq:CR_intensity}
\end{align}
where $I_{i}^{\mathrm{LIS}}(E_i)$ is the LIS, $e$ is the elementary charge, $\Phi$ is the modulation potential, $Z_{i}$ is the atomic number, $A_{i}$ is the mass number, and $m_{i}c^2$ is the rest mass energy for $i=\mathrm{p,\,\alpha}$. For the LIS $I_{i}^{\mathrm{LIS}}(E_i)$, we use version 54 of GALPROP WebRun\footnote{See \url{https://galprop.stanford.edu/webrun/}; also \citet{vladimirov+11}.}, which solves the transport equation to model the propagation of CRs through the Galaxy \citep[][]{strong_moskalenko98, strong+07, porter+17, porter+22}. 
We adopt the parameter set of Model 3 from \citet{tsuji+23}\footnote{See also the file ``galdef\_54\_0abb001i'', available online, at \url{https://github.com/tsuji703/MeV-All-Sky/tree/main/files/allsky}}, to reproduce the LIS derived from the DRELowV model in \citet{orlando18}. This parameter set was chosen to be consistent with CR measurements, Galactic diffuse gamma-ray data, and radio synchrotron data based on a CR propagation model with diffusion and reacceleration.

Figure~\ref{fig:CRspec} compares our modeled CR spectra with observational data from \textit{Voyager 1} \citep{cummings+16} and AMS-02  \citep{aguilar+15_p, aguilar+15_a}. The solid and dashed lines show CR proton and alpha particle spectra, respectively, derived from GALPROP for two cases: an unmodulated LIS (blue) and moderate solar modulation effect ($\Phi=500\,\mathrm{MV}$; red). The modulation potential is the same as in \citetalias{moskalenko_porter07}. Note that the observational data includes minor contributions from isotopes such as deuterons and $\mathrm{^{3}He}$. These CR spectral models provide the input CR intensities for our gamma-ray calculations.

Since some cosmogenic nuclides like $\mathrm{^{26}Al}$ and $\mathrm{^{40}K}$ have very long lifetimes ($\sim1\,\mathrm{Myr}$ and $\sim1.8\,\mathrm{Gyr}$, respectively), a complete treatment of gamma-ray emission from these isotopes would require modeling CR intensity variations over Gyr timescales. Given the limited information about such long-term variations, we ignore them and assume that the injected CR intensities are time-independent. Instead, we calculate time-independent
gamma-ray spectra for different solar modulation conditions, assuming the spectrum reaches a steady state after all radioactive decay lines appear. This assumption should be tested with future lunar MeV gamma-ray observations.

\section{Lunar Gamma-ray Yields and Spectrum} \label{sec:result}

Figure~\ref{fig:Gamyield} shows the gamma-ray yields obtained from {\tt Geant4} simulations for different primary CR protons and alpha particles using {\tt QGSP\_BIC}. The color maps illustrate the dependence of the gamma-ray yields $Y(E_\gamma|E_i)$ on the energy of incident CRs, $E_i$, and the resulting gamma-ray energy, $E_\gamma$. These yield maps provide key insights into the gamma-ray production mechanisms and their energy distributions across different CR interaction regimes.

Lunar gamma-ray emission arises from CR interactions at $E_{{\rm p,\,\alpha}} \gtrsim 10\,{\rm MeV\,nuc^{-1}}$ and shows characteristic features across the energy spectrum. The gamma-ray yields exhibit two distinct regions corresponding to the underlying physical processes.  In the low-energy region, $E_\gamma \lesssim 10~\mathrm{MeV}$, yields are dominated by nuclear de-excitation and radioactive decay processes. These produce sharp spectral line features arising from discrete energy level transitions, such as the prominent 6.129~MeV line from $^{16}$O de-excitation, the 1.809~MeV line from $^{26}$Al decay, and the 511~keV line from the pair annihilation. By contrast, the high-energy regime, $E_\gamma \gtrsim 100~\mathrm{MeV}$, is characterized by continuum emission driven by hadronic decay processes, dominated by $\pi^0\rightarrow\gamma\gamma$. Between these two energy regions, bremsstrahlung radiation from secondary charged particles interacting with the lunar surface material creates a broad transitional feature. The overall emission declines at $E_{\gamma}\gtrsim 1\,{\rm GeV}$, and this high-energy cutoff appears because CRs with $E_{{\rm p,\,\alpha}}>1\,\mathrm{GeV\,nuc^{-1}}$ penetrate deeper into the lunar surface layer, beyond the mean free paths of generated gamma-ray photons \citepalias[see Section 2 in][]{moskalenko_porter07}.

Compared to the contribution of primary CR protons to gamma-ray emission, alpha particles can generate gamma-rays with a lower energy per nucleon, and provide more intense yields for each physical process. However, the contribution of alpha particles to the overall gamma-ray emission is $\sim 20\%$ of that of protons when accounting for the composition of Galactic CRs (see \S\ref{subsec:method_CR}).

\begin{figure}[t!]
\centering
 \includegraphics[width=1.0\linewidth]{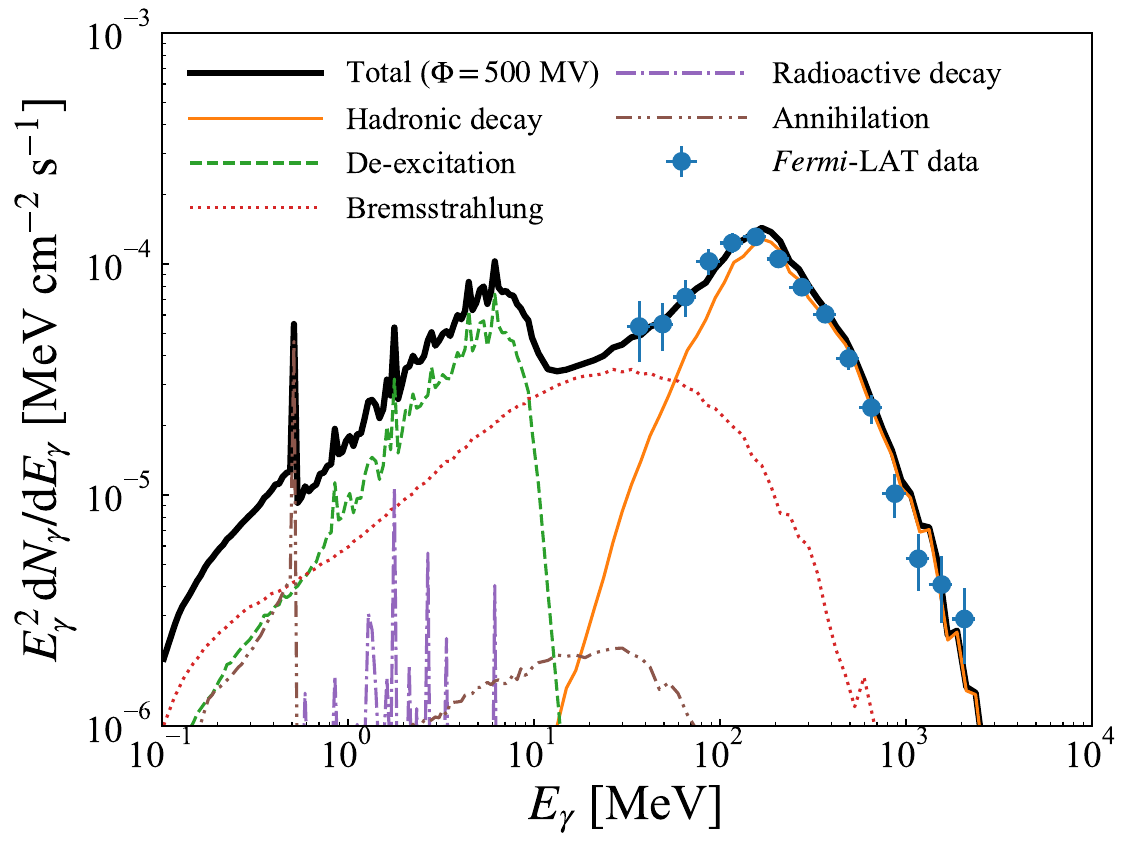}
    \caption{The lunar MeV-GeV gamma-ray spectrum along with the decomposed gamma-ray components from various physical processes. The black solid line shows the total spectrum. The green dashed line shows gamma-ray components from nuclear de-excitation, while other mechanisms are also indicated in the legend. Blue points are lunar gamma-ray data taken by \textit{Fermi}-LAT \citep[2011 May--2013 November,][]{ackermann+16}. The gamma-ray energy $E_{\gamma}$ is divided into equally spaced logarithmic scale bins: 100 bins from $0.1\,\mathrm{MeV}$ to $100\,\mathrm{MeV}$ and 60 bins for energies above $100\,\mathrm{MeV}$.}
    \label{fig:Gamspec_component}
\end{figure}

Figure \ref{fig:Gamspec_component} shows the lunar MeV-GeV gamma-ray spectrum decomposed into the contributing physical processes, along with observational data from \textit{Fermi}-LAT \citep{ackermann+16}. We obtain the expected lunar gamma-ray spectrum by combining Galactic CR intensity and gamma-ray yields. The thick solid line represents the total spectrum, while other lines indicate contributions from specific gamma-ray production mechanisms: hadronic decay (thin solid), nuclear de-excitation (dashed), bremsstrahlung (dotted), radioactive decay (dot-dashed), and electron-positron annihilation (double-dot-dashed). Other emission processes such as neutron capture, cascade, and capture at rest are not shown as they make minor contributions over the considered spectral range. Nuclear de-excitation and hadronic decay processes dominate in the $E_{\gamma} \leq 10~\mathrm{MeV}$ and $100~\mathrm{MeV} \lesssim E_{\gamma}$ ranges, respectively, forming two prominent spectral bumps.  Hadronic decay processes are able to reproduce the entire \textit{Fermi}-LAT spectrum. Bremsstrahlung emission contributes broadly, peaking between these two processes, while the intense $511~\mathrm{keV}$ electron-positron annihilation line and its associated positronium continuum are clearly visible at lower energies. Multiple gamma-ray line features also appear in addition to the $511~\mathrm{keV}$ electron-positron annihilation line. Some of these nuclear gamma-ray lines, such as those from $^{16}$O and $^{26}$Al, could exhibit fluxes above $10^{-8}\,{\rm ph\,cm^{-2}\,s^{-1}}$, making them detectable by several next-generation MeV gamma-ray missions (see \S\ref{subsec:disc_lunarMeV_prospect}).

\section{Discussion} \label{sec:discussion}

\subsection{Comparison with Previous Studies} \label{subsec:disc_comaprison}

\begin{figure}
    \plotone{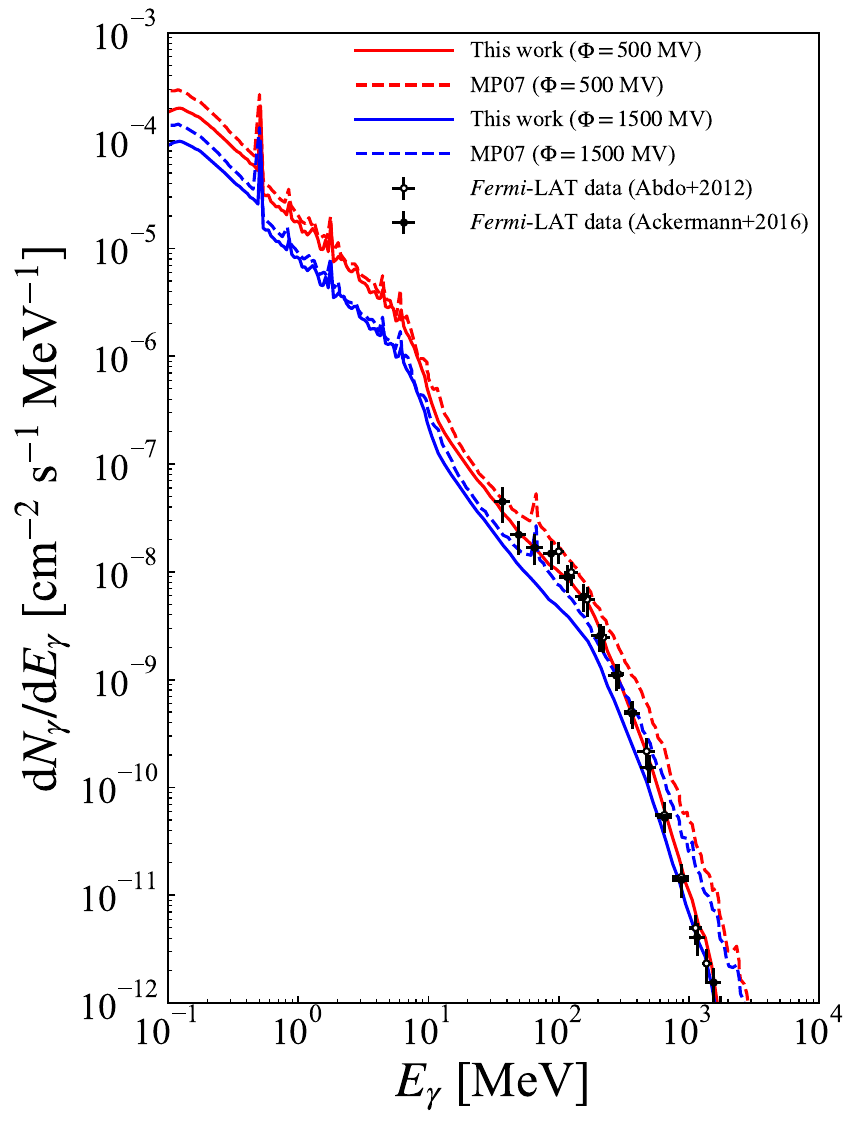}
    \caption{The lunar MeV-GeV gamma-ray fluxes for this work (solid line) in comparison to the results obtained by \citetalias{moskalenko_porter07} (dashed line). Open and filled points are gamma-ray data for two years \citep[2008 August--2010 August,][]{abdo+12} and seven years \citep[2008 August--2015 June,][]{ackermann+16} of observations by the \textit{Fermi}-LAT, respectively.}
    \label{fig:Gam_spec_wMP07}
\end{figure}

We compare our results with \citetalias{moskalenko_porter07}, a  pioneering study in modeling lunar gamma-ray emission, 
to evaluate how advances in simulations have improved our capability to model the underlying physical processes. \citetalias{moskalenko_porter07} used version 8.2.0 of the {\tt Geant4} toolkit, being the latest version available at the time, to simulate the lunar gamma-ray albedo. To aid comparison, we adopt an identical lunar surface parameters to \citetalias{moskalenko_porter07} while using updated physical models in our simulations. To ensure a meaningful comparison, we adopt an identical solar-terrestrial CR spectrum as \citetalias{moskalenko_porter07}. 

Figure~\ref{fig:Gam_spec_wMP07} compares the total gamma-ray spectra predicted by \citetalias{moskalenko_porter07} with our model using {\tt QGSP\_BIC} for solar modulation potentials of $\Phi=500$ MV (red) and $1500$~MV (blue), in units of ${\rm d}N_\gamma/{\rm d}E_\gamma$. A notable difference appears at $67.5~\mathrm{MeV}$, where \citetalias{moskalenko_porter07} predicted a narrow $\pi^0$-decay line and  interpreted it as a potentially unique astrophysical feature. This feature is absent in our simulations, which instead show better agreement with the seven-year \textit{Fermi}-LAT observations \citep{ackermann+16}.

The comparison also reveals differences in the high-energy continuum. \citet{abdo+12} pointed out that the \textit{Fermi}-LAT spectrum at $\gtrsim100\,{\rm MeV}$ is softer than predicted by \citetalias{moskalenko_porter07}. It has been discussed that this discrepancy might arise if the observed emission originates primarily from the inner lunar disk (diameter of $0.48\deg$), with a negligible limb contribution. Our model, however, naturally reproduces  the \textit{Fermi}-LAT observations \citep{abdo+12, ackermann+16} when considering the Moon's full angular extent of $0.52\deg$ (cf., $2\times(R_{\leftmoon}/a_{\rm EM})\times(180/\pi)\approx 0.52\deg$). This agreement suggests that the observed lunar gamma-ray emission originates from the entire lunar disk rather than being confined to specific regions.

\subsection{Comparison of Physics Models} \label{subsec:disc_InteractionModel}

\begin{figure*}[t!]
\centering
    \gridline{\fig{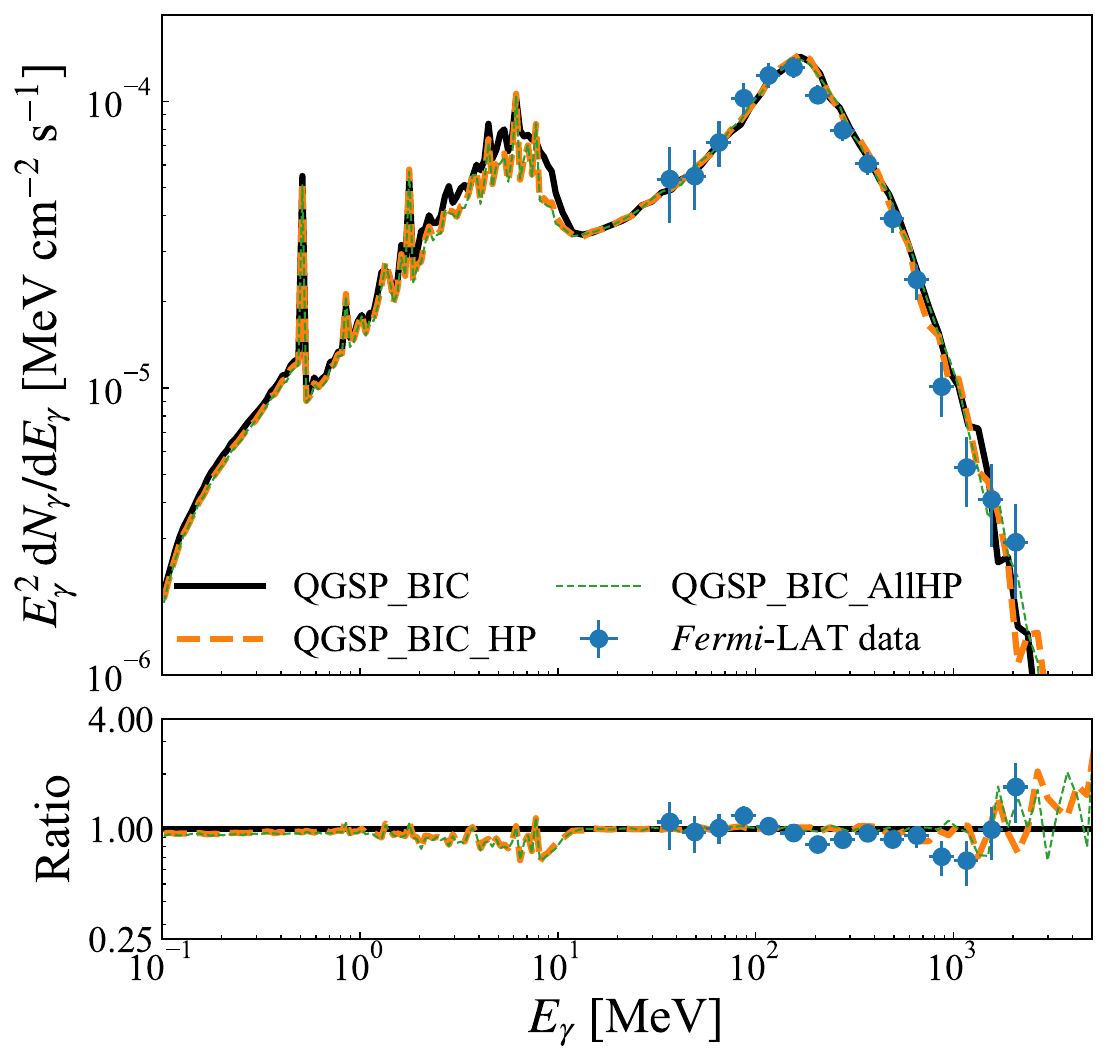}{0.45\textwidth}{(a) Comparison between our baseline model, \texttt{QGSP\_BIC\_HP} and \texttt{QGSP\_BIC\_AllHP}}
              \fig{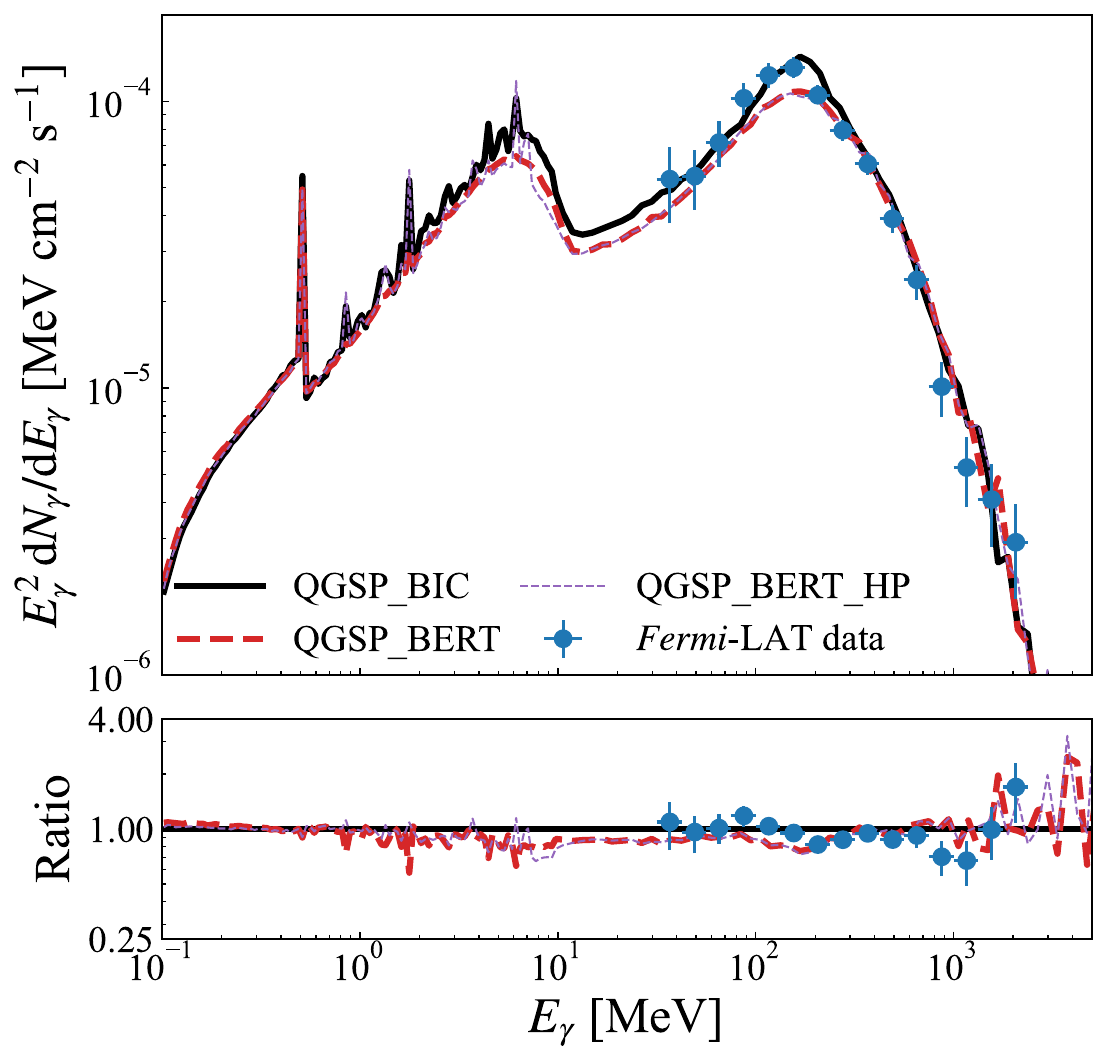}{0.45\textwidth}{(b) Comparison between our baseline model, \texttt{QGSP\_BERT} and \texttt{QGSP\_BERT\_HP}}}
    \gridline{\fig{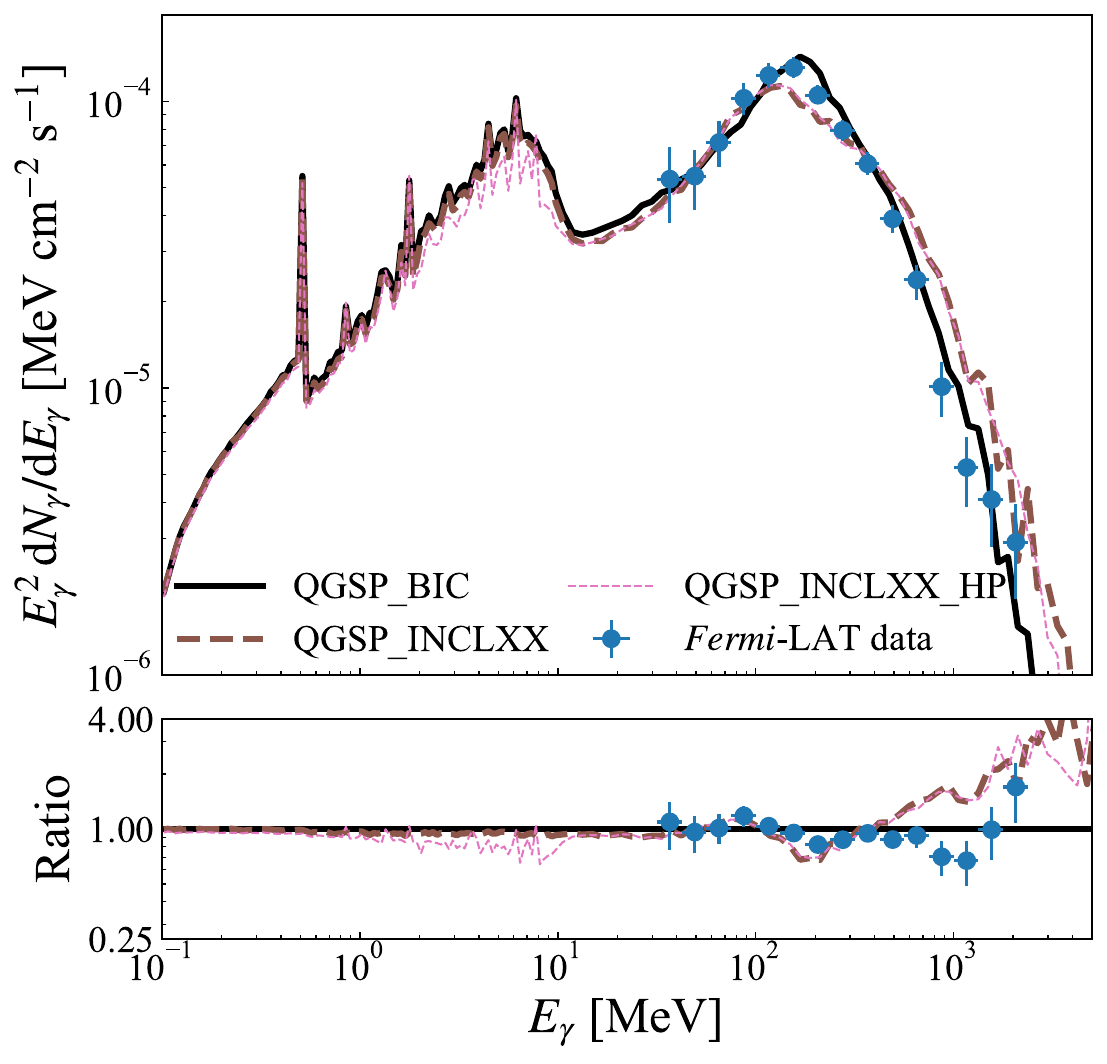}{0.45\textwidth}{(c) Comparison between our baseline model, \texttt{QGSP\_INCLXX} and \texttt{QGSP\_INCLXX\_HP}}
              \phantom{\fig{Fgam_inclxx_comparison.pdf}{0.45\textwidth}{}}}
\caption{Comparison of lunar gamma-ray spectra computed using different Physics Lists. All panels show \texttt{QGSP\_BIC} (baseline) results in solid lines, compared with (a) \texttt{QGSP\_BIC\_HP} (dashed) and \texttt{QGSP\_BIC\_AllHP} (dotted), (b) \texttt{QGSP\_BERT} (dashed) and \texttt{QGSP\_BERT\_HP} (dotted), and (c) \texttt{QGSP\_INCLXX} (dashed) and \texttt{QGSP\_INCLXX\_HP} (dotted). The bottom panels show the corresponding ratios to the baseline results. Data points are the same as in Fig.~\ref{fig:Gamspec_component}.}
    \label{fig:Gamspec_comp_all}
\end{figure*}

To understand how different hadronic interaction models affect our predictions, we compare the results from different choices of Physics Lists with our baseline \texttt{QGSP\_BIC} model.
The most significant variations appear in the predicted gamma-ray spectra, shown in Figure~\ref{fig:Gamspec_comp_all}. 

Among the Binary Cascade (BIC) family of models, \texttt{QGSP\_BIC\_HP} and \texttt{QGSP\_BIC\_AllHP} predict slightly enhanced line fluxes in the $0.1~\mathrm{MeV} \leq E_{\gamma} \leq 10~\mathrm{MeV}$ range compared to the baseline \texttt{QGSP\_BIC}. This is primarily due to increased neutron inelastic scattering and capture processes. Most notably, the $^{16}$O de-excitation line is enhanced by a factor of about 12 relative to the baseline, while the differences in other prominent gamma-ray lines remain modest ($\lesssim 10\%$). For instance, for the 511 keV electron-positron annihilation line, {\tt QGSP\_BIC\_HP} and {\tt QGSP\_BIC\_AllHP} show reductions of $\approx 5\%$ and $8\%$, respectively.

The Bertini cascade models (\texttt{QGSP\_BERT} and \texttt{QGSP\_BERT\_HP}) show more distinct behavior, producing fewer nuclear de-excitation lines and radioactive decay signatures. This is due to their simpler precompound and de-excitation schemes. 
For the 511 keV line, both models predict smaller fluxes by $\approx 13\%$ compared to {\tt QGSP\_BIC}.
These models also underestimate the flux at $E_{\gamma} \sim 100~\mathrm{MeV}$ compared to \textit{Fermi}-LAT observations, though they match the data well at higher energies. This discrepancy arises from a reduced production of cascade-induced and hadronic-decay gamma-rays in this energy range.

The INCL++ cascade models (\texttt{QGSP\_INCLXX} and \texttt{QGSP\_INCLXX\_HP}) show a third pattern, under-producing flux at $E_{\gamma} \sim 100~\mathrm{MeV}$ while overestimating it at higher energies. While their MeV gamma-ray predictions are comparable to \texttt{QGSP\_BIC}, e.g., a reduction of $\approx 3\%$ for {\tt QGSP\_INCLXX} and $5\%$ for {\tt QGSP\_INCLXX\_HP} to the baseline in the 511 keV line flux, they generate cascade-induced and hadronic-decay gamma-rays with notably different energy distributions in the 100 MeV to GeV range.

Future observations with next-generation MeV gamma-ray instruments will allow hadronic 
models to be stringently tested.  By precisely measuring nuclear de-excitation lines and radioactive decay signatures that are sensitive to low-energy hadronic processes, future data 
will allow a confident determination of the accuracy of the underlying cross-sections, and allow a substantial refinement of hadronic interaction models.

\subsection{Comparison with Observations by Kaguya} \label{subsec:Kaguya}

The \textit{Kaguya} (SELenological and ENgineering Explorer, SELENE) mission, launched in 2007, was a Japanese lunar orbiter designed to study the origin and evolution of the Moon. Its gamma-ray spectrometer (GRS) provided comprehensive measurements of the lunar surface composition through detection of characteristic gamma-ray lines \citep[e.g.,][]{yamashita+10, yamashita+12, naito+18}. Notably, the \textit{Kaguya} GRS offers superior energy resolution compared to other lunar orbiter GRSs \citep[e.g., see Table 1 in][]{yamashita+15}.
While the GRS was primarily designed for elemental abundance mapping, these observations offer important benchmarks for validating our theoretical predictions of lunar gamma-ray emission.

Comprehensive comparison across multiple isotopes remains challenging due to differences in observational conditions and background treatments between our simulations and actual measurements. For instance, the \textit{Kaguya} GRS itself and the spacecraft body could significantly affect observations of Mg, Al, and Si gamma-ray lines because Mg and Al are contained in the spacecraft structure, the gamma-ray detector is enclosed by an Al canister \citep[e.g.,][]{kobayashi+10, kobayashi+13}, and Si is produced in decay channels of Al following a CR interaction. 
Measurements of the long-lived $^{26}$Al decay line ($1.809$ MeV)\footnote{$^{26}{\rm Al}\rightarrow ^{26}{\rm Mg^*}\rightarrow \gamma\,(1.809\,{\rm MeV})+^{26}{\rm Mg}$}, which is one of the most intriguing targets for the study of the history of local MeV CRs, can therefore be prone to contamination resulting from the direct production of $^{26}{\rm Mg^*}$ by CR colliding with $^{27}$Al in the GRS itself. 
Moreover, to rigorously compare our theoretical results with these observations, the spatial response function (SRF) of the GRS should be considered, which determines the directional sensitivities. Once the SRF effect is disentangled from the observed count rates, a comparable gamma-ray line flux can be derived. Such detailed analyses for the \textit{Kaguya} GRS data, not yet publicly available in the literature or a database, are beyond the scope of this paper and reserved for our future work.

Future lunar missions with improved gamma-ray spectroscopy capabilities could provide more stringent model tests. In particular, measurements of both prompt de-excitation lines and long-lived radioactive decay signatures would validate different aspects of our calculations, from immediate CR interaction processes to the long-term accumulation of cosmogenic nuclides in the lunar surface.

\subsection{Sporadic CR Bombardment of the Moon} \label{subsec:disc_otherInjectionModels}

While our primary analysis focuses on steady-state gamma-ray emission induced by Galactic CRs, sporadic enhancements in the CR flux can significantly contribute to lunar gamma-ray variability. Here we examine two such sources: solar energetic particles (SEPs) and anomalous cosmic rays (ACRs).

\subsubsection{SEPs} \label{subsubsec:disc_SEPs}

SEPs are non-thermal particles accelerated by solar activity, with energies ranging from a few keV to several GeV \citep[e.g.,][]{desai_giacalone16}. These particles are produced through two distinct mechanisms: (i) impulsive events driven by magnetic reconnection in solar flares \citep[e.g.,][]{reames99, knizhnik+11}, and (ii) gradual events generated by coronal mass ejection (CME)-driven interplanetary shocks \citep[e.g.,][]{reames99, desai_giacalone16}. Although these events differ in duration and particle composition, both produce accelerated electrons, protons, and heavier nuclei. Gradual events are particularly relevant for lunar gamma-ray production as they generate protons with energies $E_{\rm p}\gtrsim10\,{\rm MeV\,nuc^{-1}}$. In extreme cases, these particles can reach GeV energies, allowing them to penetrate Earth's atmosphere and reach the ground. Such events are known as ground-level events (GLEs; \citealt{reames_ng_10}).

To quantify the potential impact of SEPs, we simulated the lunar gamma-ray emission resulting from an intense GLE, following the characteristics of the event of 2003 October 28, with a particle spectrum following that reported by \citet{reames_ng_10}. Our calculations show that during such an event ($\sim$ day scale), the gamma-ray flux would be enhanced by approximately two orders of magnitude compared to our baseline Galactic CR-induced emission with $\Phi=500\,{\rm MV}$.

\subsubsection{ACRs} \label{subsubsec:disc_ACRs}

ACRs constitute a distinct population of low-energy ($<100\,\mathrm{MeV\,nuc^{-1}}$) particles accelerated within the heliosphere \citep[for a recent review, see][]{giacalone+22}. They were first identified through an unexpected enhancement in the CR $\mathrm{^{4}He}$ spectrum below $\lesssim50\,\mathrm{MeV\,nuc^{-1}}$ \citep{garcia-munoz+73, garcia-munoz+75}, with similar features subsequently observed in H, N, O, and Ne spectra \citep[e.g.,][]{cummings_stone_07, giacalone+22}.

While their acceleration mechanism remains under debate, ACRs are widely believed to originate from interstellar pickup ions that undergo diffusive shock acceleration at the solar wind termination shock \citep{fisk+74, giacalone+22}. In this work, we consider ACRs as a population encompassing both traditional ACRs and termination shock particles, which represent the high and low-energy components, respectively.

\textit{Voyager} measurements have demonstrated that ACR fluxes can exceed those of Galactic CRs by more than an order of magnitude at $>1\,{\rm au}$ \citep[e.g.,][]{cummings+02, krimigis+03, mcdonald+03, decker+05, stone+05, stone+08, stone+13}. Using ACR spectra observed by \textit{Voyager~1} near the termination shock \citep{stone+05}\footnote{The ACR spectral data in \citet{stone+05} were obtained by \textit{Voyager 1} in 2005 at $\sim 95\,{\rm au}$, near the termination shock region. At energies $\gtrsim 100\,{\rm MeV\,nuc^{-1}}$, the observed fluxes were nearly identical to those of Galactic CRs with $\Phi=500\,{\rm MV}$.}, we estimate that ACRs could enhance the lunar MeV gamma-ray emission by a few percent relative to the Galactic CR contribution. This can be regarded as an upper limit of the ACR contribution. The enhancement varies between physics models but generally accounts for 5-10\% of the MeV de-excitation continuum and nuclear line components. This variation reflects the different spectral shapes and normalizations of ACRs compared to Galactic CRs, resulting in a modest but potentially detectable modification of the lunar MeV gamma-ray spectrum.

\begin{figure}[t!]
\centering
    \includegraphics[width=1.0\linewidth]{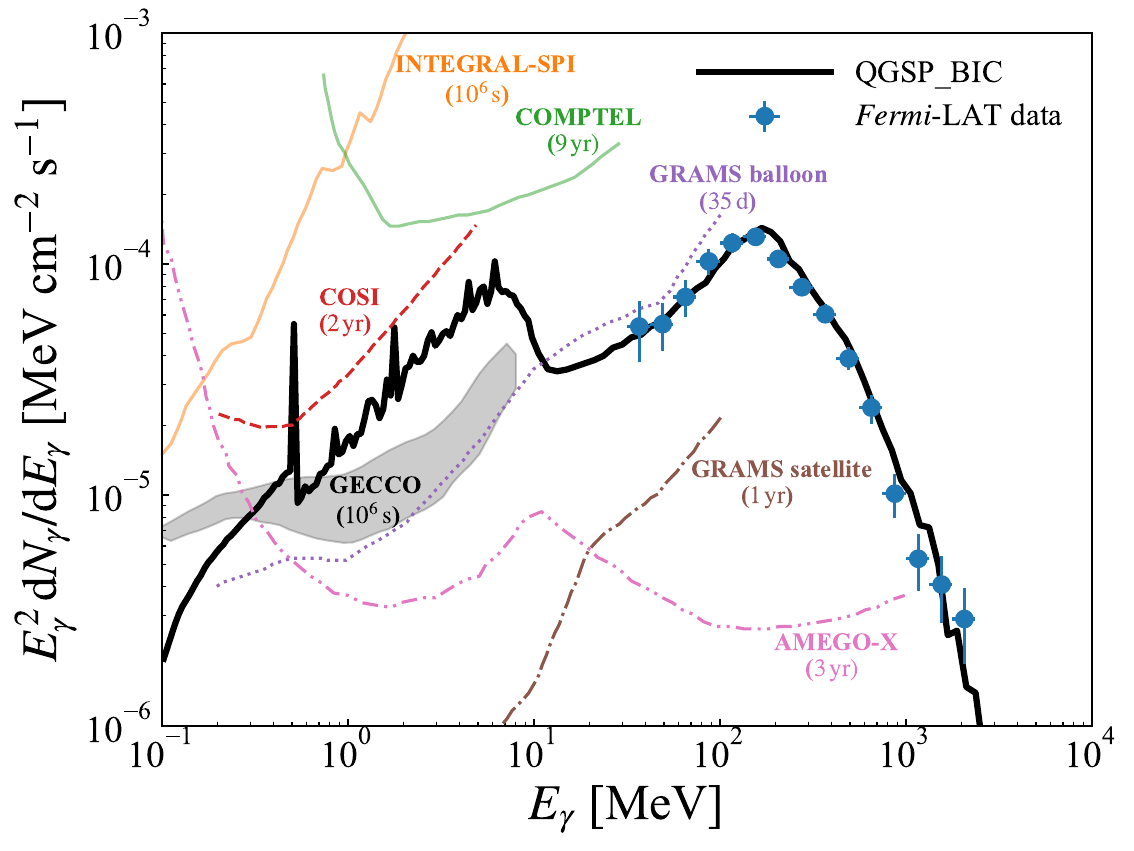}
    \caption{Same as Fig.~\ref{fig:Gamspec_component} computed by \texttt{QGSP\_BIC}, but showing  the comparison with continuum gamma-ray sensitivities of various instruments, including COMPTEL \citep{schonfelder+84, takahashi+13}, INTEGRAL-SPI \citep{vedrenne+03, deangelis+17}, COSI \citep{tomsick+23}, GRAMS \citep{aramaki+20}, AMEGO-X \citep{fleischhack+22}, and GECCO \citep{orlando+22}.}
    \label{fig:Gamspec_instruments}
\end{figure}

\subsection{Prospects for Future MeV Gamma-ray Missions} \label{subsec:disc_lunarMeV_prospect}
Figure~\ref{fig:Gamspec_instruments} presents our predicted lunar gamma-ray spectrum using the \texttt{QGSP\_BIC} model alongside sensitivity thresholds of current and future MeV gamma-ray instruments. The predicted spectrum falls below the detection capabilities of earlier instruments such as COMPTEL \citep{schonfelder+84} and INTEGRAL-SPI \citep{vedrenne+03}, consistent with their non-detections in the MeV band. However, next-generation missions including COSI \citep{tomsick+23}, AMEGO-X \citep{fleischhack+22}, GECCO \citep{orlando+22}, e-ASTROGAM \citep{deangelis+18}, GammaTPC \citep{shutt+25}, and GRAMS \citep{aramaki+20} should be capable of detecting lunar gamma-ray emission.

Based on the predicted sensitivities shown in Figure~\ref{fig:Gamspec_instruments}, instruments such as GECCO, GRAMS, and AMEGO-X will be able to detect the lunar MeV continuum, providing insights into the solar-terrestrial CR environment down to energies of $10\,{\rm MeV\,nuc^{-1}}$. During GLEs (see \S\ref{subsubsec:disc_SEPs}), the gamma-ray flux could increase by up to two orders of magnitude. Assuming instrumental sensitivity scales as $\propto T_{\rm obs}^{-1/2}$, where $T_{\rm obs}$ is the observation time, COSI could detect GLE-induced gamma-ray emission during these enhanced periods.

Our analysis reveals multiple discrete gamma-ray lines in the lunar spectrum (Figure~\ref{fig:Gamspec_component}), with fluxes above $10^{-8}~\mathrm{ph\,cm^{-2}\,s^{-1}}$ listed in Table~\ref{tab:MeV_line_list}. These lines originate from radioactive decay, nuclear de-excitation, and neutron capture processes. The line identifications are based on compilations by \citet{ramaty+79} and \citet{evans+06}, updated with current nuclear data from ENSDF\footnote{\url{https://www.nndc.bnl.gov/ensdf/}} and validated against Martian gamma-ray observations \citep{evans+06}. We evaluate line fluxes by binning gamma-ray energies in $1~\mathrm{keV}$ intervals and calculating $\mathrm{d}E_{\gamma} \times \mathrm{d}N_{\gamma}/\mathrm{d}E_{\gamma}$ within each bin.

\begin{figure}
\plotone{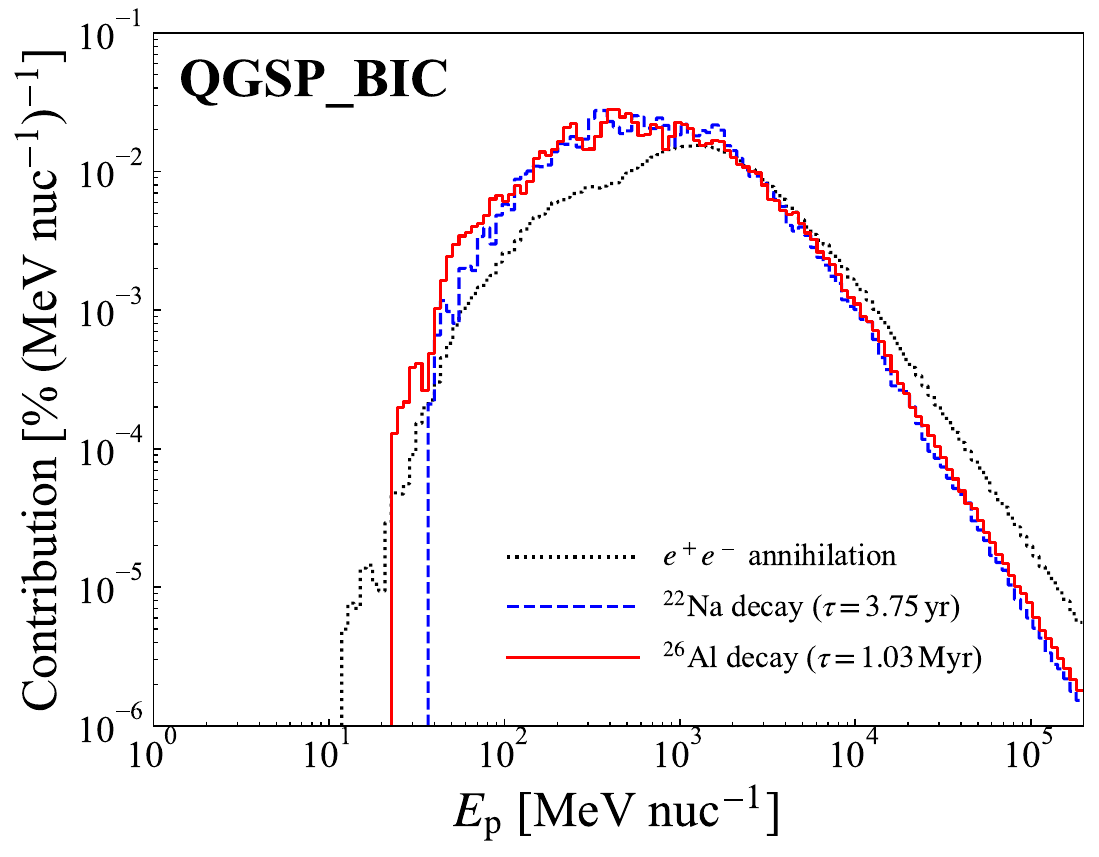}
    \caption{The contribution of Galactic CR protons to the production of $e^+e^-,\,{\rm ^{22}Na,\,\text{and }^{26}Al}$ gamma-ray lines. The calculations are conducted with the Galactic CR model of \citet{orlando18}, and {\tt QGSP\_BIC}. $\tau$ in the legend is the lifetime of the radioactive isotope.}
    \label{fig:CRp_contrib_to_MeVline}
\end{figure}

While detecting individual lines remains challenging even for next-generation missions, several future missions could achieve sufficient sensitivity to observe key gamma-ray lines (Table~\ref{tab:MeV_line_list}). These lines offer unique temporal probes of CR activity across different timescales. 
The $0.511\,{\rm MeV}$ $e^+e^-$ annihilation line is one of the most promising targets. The predicted pair annihilation flux is $4.0\times10^{-6}\,{\rm ph\,cm^{-2}\,s^{-1}}$, while the detection limit of the COSI 2-year survey sensitivity requirement is expected to be $1.2\times10^{-5}\,{\rm ph\,cm^{-2}\,s^{-1}}$ \citep[][]{tomsick+23}. We note that the solar modulation condition (i.e., $\Phi$) changes the expected flux: $8.5\times10^{-6}$ and $1.9\times10^{-6}\,{\rm ph\,cm^{-2}\,s^{-1}}$ for $\Phi=0$ and $1500\,{\rm MV}$, respectively\footnote{The predicted flux for $\Phi=0\,{\rm MV}$ shows $\approx 2.8$ times lower than that by \citetalias{moskalenko_porter07} \citep[the value indicated in \S5 in][]{moskalenko+08}. This difference is substantially larger than the systematic error arising from the {\tt Geant4} Physics Lists ($\lesssim13\%$; see \S\ref{subsec:disc_InteractionModel}).}. This gamma-ray line provides a probe of current CR interactions. Nuclear de-excitation lines, such as the prominent $1.779~\mathrm{MeV}$ line from $\mathrm{^{28}Si}$ ($\tau = 6.85\times10^{-13}~\mathrm{s}$), also serve as direct tracers of current CR activity due to their short lifetimes. This particular line is enhanced by the high silicon content (45\% $\mathrm{SiO_2}$ by weight) in the lunar surface material (see \S\ref{subsubsec:method_LunarModel}). On the other hand, radioactive decay lines trace various historical periods: the $1.809\,{\rm MeV}$ line from $^{26}$Al ($\tau \sim 1\,{\rm Myr}$) probes long-term CR history, and isotopes like $^{54}$Mn and $^{22}$Na (with lifetimes of years to decades) reveal more recent variations. Figure~\ref{fig:CRp_contrib_to_MeVline} shows that these lines are predominantly produced by CRs with energies $\lesssim 1\,{\rm GeV\,nuc^{-1}}$, extending our observational window below the energy range accessible with \textit{Fermi}-LAT.
\pagebreak
\startlongtable
\begin{deluxetable*}{cccc}
\centering
\tablecaption{List of MeV gamma-ray lines}
\label{tab:MeV_line_list}
\tablehead{
\colhead{Energy [$\mathrm{MeV}$]} & \colhead{Flux [$\mathrm{ph \,cm^{-2} \,s^{-1}}$]} & \colhead{Nuclear/Particle Process} & \colhead{Lifetime}
}
\startdata
$0.417$ & $2.5 \times 10^{-8}$ & $\mathrm{^{26}Al^* \rightarrow ^{26}Al}$ & $1.73\times10^{-9}\,{\rm s}$ \\
$0.511$ & $4.0 \times 10^{-6}$ & $e^++e^-\rightarrow \gamma+\gamma$ & --- \\
$0.835$ & $4.2 \times 10^{-8}$ & $\mathrm{^{54}Mn\rightarrow ^{54}Cr^*\rightarrow ^{54}Cr}$ & $\mathrm{1.23\,yr}$ \\
$0.844$ & $1.3 \times 10^{-8}$ & $\mathrm{^{27}Al^* \rightarrow ^{27}Al}$ & $\mathrm{5.0\times10^{-11}\,s}$ \\
$0.847$ & $1.1 \times 10^{-8}$ & $\mathrm{^{56}Fe^* \rightarrow ^{56}Fe}$ & $\mathrm{8.8\times10^{-12}\,s}$ \\
$0.984$ & $1.5 \times 10^{-8}$ & $\mathrm{^{48}Ti^*\rightarrow ^{48}Ti}$ & $\mathrm{6.5\times10^{-12}\,s}$ \\
$1.275$ & $6.5 \times 10^{-8}$ &  $\mathrm{^{22}Na\rightarrow ^{22}Ne^*\rightarrow ^{22}Ne}$ & $\mathrm{3.75\,yr}$ \\
$1.369$ & $6.5 \times 10^{-8}$ & $\mathrm{^{24}Mg^* \rightarrow ^{24}Mg}$ & $\mathrm{1.96\times10^{-12}\,s}$ \\
$1.634$ & $1.3 \times 10^{-8}$ & $\mathrm{^{20}Ne^* \rightarrow ^{20}Ne}$ & $\mathrm{1.1\times10^{-12}\,s}$ \\
$1.779$ & $1.2 \times 10^{-7}$ & $\mathrm{^{28}Si^* \rightarrow ^{28}Si}$ & $\mathrm{6.85\times10^{-13}\,s}$ \\
$1.809$ & $1.4 \times 10^{-7}$ & $\mathrm{^{26}Al \rightarrow ^{26}Mg^*\rightarrow ^{26}Mg}$ & $\mathrm{1.03\,Myr}$ \\
$2.313$ & $1.8 \times 10^{-8}$ & $\mathrm{^{14}N^* \rightarrow ^{14}N}$ & $\mathrm{9.8\times10^{-14}\,s}$ \\
$6.129$ & $3.0 \times 10^{-8}$ & $\mathrm{^{16}O^* \rightarrow ^{16}O}$ & $\mathrm{2.65\times10^{-11}\,s}$ \\
\enddata
\tablecomments{This table includes the fluxes, decay chains, and lifetimes for gamma-ray lines predicted by the \texttt{QGSP\_BIC} model.}
\end{deluxetable*}

\section{Conclusions} \label{sec:conclusion}

In this study, we investigate the MeV gamma-ray emission from the Moon using state-of-the-art hadronic interaction models and detailed lunar surface compositions. Our comprehensive analysis establishes lunar gamma-ray observations as a valuable diagnostic tool for studying CR interactions in the heliosphere. 

Our simulations predict a characteristic two-bump spectrum arising from nuclear de-excitation ($\lesssim 10~\mathrm{MeV}$) and hadronic decay ($\gtrsim 100~\mathrm{MeV}$) processes, with some variations between different physics models - particularly in the treatment of low-energy nuclear interactions. This sensitivity to model choice highlights the importance of cross-section uncertainties in the MeV regime and underscores the potential value of lunar gamma-ray observations for constraining nuclear physics models.

Our analysis reveals multiple gamma-ray line features that could serve as powerful probes of CR interactions across different timescales. The prominent $1.779~\mathrm{MeV}$ line from $^{28}\mathrm{Si}$ de-excitation provides an immediate tracer of current CR bombardment, enhanced by the silicon-rich lunar surface composition. The $1.809~\mathrm{MeV}$ line from long-lived $^{26}\mathrm{Al}$ ($\tau \sim 1~\mathrm{Myr}$) offers insights into historical CR activities, while lines from intermediate-lifetime isotopes like $^{54}\mathrm{Mn}$ and $^{22}\mathrm{Na}$ probe more recent variations. These features are predominantly produced by CRs with energies $\lesssim 1~\mathrm{GeV\,nuc^{-1}}$, extending our observational window below the energy range accessible with current instruments.

Next-generation MeV gamma-ray missions such as COSI, GRAMS, AMEGO-X, GECCO, e-ASTROGAM, and GammaTPC should be capable of detecting the lunar gamma-ray continuum and potentially specific nuclear lines, providing crucial insights into the solar-terrestrial CR environment down to energies of $10~\mathrm{MeV\,nuc^{-1}}$. The combination of continuum measurements and line spectroscopy promises to open a new window into both the current state of CRs in our solar neighborhood and its long-term evolution.

\section*{Acknowledgements} 
We thank an anonymous referee for their constructive report. We also would like to thank Mario Nicola Mazziotta and Thomas Siegert for their fruitful comments.
TF is supported by JST SPRING, Grant Number JPMJSP2138.
ERO is an international research fellow under the Postdoctoral Fellowship of the Japan Society for the Promotion of Science (JSPS), supported by JSPS KAKENHI Grant Number JP22F22327. ERO also acknowledges support from the Special Postdoctoral Researcher Program for junior scientists at RIKEN, where some of this work was completed. 
YI is supported by NAOJ ALMA Scientific Research Grant Number 2021-17A; JSPS KAKENHI Grant Number JP18H05458, JP19K14772, and JP22K18277; and World Premier International Research Center Initiative (WPI), MEXT, Japan. 
KN and KO are supported by ISEE International Joint Research Program Grant Number 24H00244. HY is supported by JSPS KAKENHI Grant Number 23K13136.

%






\bibliography{sample631}{}

\begin{thebibliography}{}
\expandafter\ifx\csname natexlab\endcsname\relax\def\natexlab#1{#1}\fi
\providecommand{\url}[1]{\href{#1}{#1}}
\providecommand{\dodoi}[1]{doi:~\href{http://doi.org/#1}{\nolinkurl{#1}}}
\providecommand{\doeprint}[1]{\href{http://ascl.net/#1}{\nolinkurl{http://ascl.net/#1}}}
\providecommand{\doarXiv}[1]{\href{https://arxiv.org/abs/#1}{\nolinkurl{https://arxiv.org/abs/#1}}}

\bibitem[{{Aad} {et~al.}(2012){Aad}, {Abajyan}, {Abbott}, {Abdallah}, {Abdel Khalek}, {Abdelalim}, {Abdinov}, {Aben}, {Abi}, {Abolins}, {Abouzeid}, {Abramowicz}, {Abreu}, {Acharya}, {Adamczyk}, {Adams}, {Addy}, {Adelman}, {Adomeit}, {Adragna}, {Adye}, {Aefsky}, {Aguilar-Saavedra}, {Agustoni}, {Aharrouche}, {Ahlen}, {Ahles}, {Ahmad}, {Ahsan}, {Aielli}, {Akdogan}, {{\r{A}}kesson}, {Akimoto}, {Akimov}, {Alam}, {Alam}, {Albert}, {Albrand}, {Aleksa}, {Aleksandrov}, {Alessandria}, {Alexa}, {Alexander}, {Alexandre}, {Alexopoulos}, {Alhroob}, {Aliev}, {Alimonti}, {Alison}, {Allbrooke}, {Allport}, {Allwood-Spiers}, {Almond}, {Aloisio}, {Alon}, {Alonso}, {Alonso}, {Altheimer}, {Alvarez Gonzalez}, {Alviggi}, {Amako}, {Amelung}, {Ammosov}, {Amor Dos Santos}, {Amorim}, {Amram}, {Anastopoulos}, {Ancu}, {Andari}, {Andeen}, {Anders}, {Anders}, {Anderson}, {Andreazza}, {Andrei}, {Andrieux}, {Anduaga}, {Angelidakis}, {Anger}, {Angerami}, {Anghinolfi}, {Anisenkov}, {Anjos}, {Annovi}, {Antonaki}, {Antonelli}, {Antonov}, {Antos},
  {Anulli}, {Aoki}, {Aoun}, {Aperio Bella}, {Apolle}, {Arabidze}, {Aracena}, {Arai}, {Arce}, {Arfaoui}, {Arguin}, {Arik}, {Arik}, {Armbruster}, {Arnaez}, {Arnal}, {Arnault}, {Artamonov}, {Artoni}, {Arutinov}, {Asai}, {Ask}, {{\r{A}}sman}, {Asquith}, {Assamagan}, {Astbury}, {Atkinson}, {Aubert}, {Auge}, {Augsten}, {Aurousseau}, {Avolio}, {Avramidou}, {Axen}, {Azuelos}, {Azuma}, {Baak}, {Baccaglioni}, {Bacci}, {Bach}, {Bachacou}, {Bachas}, {Backes}, {Backhaus}, {Backus Mayes}, {Badescu}, {Bagnaia}, {Bahinipati}, {Bai}, {Bailey}, {Bain}, {Baines}, {Baker}, {Baker}, {Baker}, {Balek}, {Banas}, {Banerjee}, {Banerjee}, {Banfi}, {Bangert}, {Bansal}, {Bansil}, {Barak}, {Baranov}, {Barbaro Galtieri}, {Barber}, {Barberio}, {Barberis}, {Barbero}, {Bardin}, {Barillari}, {Barisonzi}, {Barklow}, {Barlow}, {Barnett}, {Barnett}, {Baroncelli}, {Barone}, {Barr}, {Barreiro}, {Barreiro Guimar{\~a}es da Costa}, {Barrillon}, {Bartoldus}, {Barton}, {Bartsch}, {Basye}, {Bates}, {Batkova}, {Batley}, {Battaglia}, {Battistin}, {Bauer},
  {Bawa}, {Beale}, {Beau}, {Beauchemin}, {Beccherle}, {Bechtle}, {Beck}, {Becker}, {Becker}, {Beckingham}, {Becks}, {Beddall}, {Beddall}, {Bedikian}, {Bednyakov}, {Bee}, {Beemster}, {Begel}, \& {Behar Harpaz}}]{aad+12}
{Aad}, G., {Abajyan}, T., {Abbott}, B., {et~al.} 2012, Physics Letters B, 716, 1, \dodoi{10.1016/j.physletb.2012.08.020}

\bibitem[{{Abdo} {et~al.}(2012){Abdo}, {Ackermann}, {Ajello}, {Atwoo}, {Baldini}, {Ballet}, {Barbiellini}, {Bastieri}, {Bechtol}, {Bellazzini}, {Berenji}, {Blandford}, {Bonamente}, {Borgland}, {Bottacini}, {Bouvier}, {Bregeon}, {Brigida}, {Bruel}, {Buehler}, {Buson}, {Caliandro}, {Cameron}, {Caraveo}, {Casandjian}, {Cecchi}, {Charles}, {Chekhtman}, {Chiang}, {Ciprini}, {Claus}, {Cohen-Tanugi}, {Conrad}, {Cutini}, {D'Ammando}, {de Angelis}, {de Palma}, {Dermer}, {Digel}, {Silva}, {Drell}, {Drlica-Wagner}, {Dubois}, {Favuzzi}, {Fegan}, {Focke}, {Fortin}, {Fukazawa}, {Funk}, {Fusco}, {Gargano}, {Gehrels}, {Germani}, {Giglietto}, {Giommi}, {Giordano}, {Giroletti}, {Glanzman}, {Godfrey}, {Gomez-Vargas}, {Grenier}, {Grove}, {Guiriec}, {Hadasch}, {Hays}, {Hill}, {Horan}, {Hou}, {Hughes}, {Iafrate}, {Jackson}, {J{\'o}hannesson}, {Johnson}, {Kamae}, {Katagiri}, {Kataoka}, {Kn{\"o}dlseder}, {Kuss}, {Lande}, {Larsson}, {Latronico}, {Lemoine-Goumard}, {Longo}, {Loparco}, {Lott}, {Lovellette}, {Lubrano}, {Mazziotta},
  {McEnery}, {Mehault}, {Michelson}, {Mitthumsiri}, {Mizuno}, {Moiseev}, {Monte}, {Monzani}, {Morselli}, {Moskalenko}, {Murgia}, {Naumann-Godo}, {Nolan}, {Norris}, {Nuss}, {Ohno}, {Ohsugi}, {Okumura}, {Omodei}, {Orienti}, {Orlando}, {Ormes}, {Ozaki}, {Paneque}, {Panetta}, {Parent}, {Pesce-Rollins}, {Pierbattista}, {Piron}, {Pivato}, {Poon}, {Porter}, {Prokhorov}, {Rain{\`o}}, {Rando}, {Razzano}, {Razzaque}, {Reimer}, {Reimer}, {Reposeur}, {Rochester}, {Roth}, {Sadrozinski}, {Sanchez}, {Sbarra}, {Schalk}, {Sgr{\`o}}, {Share}, {Siskind}, {Spandre}, {Spinelli}, {Stawarz}, {Takahashi}, {Tanaka}, {Thayer}, {Thayer}, {Thompson}, {Tibaldo}, {Tinivella}, {Torres}, {Tosti}, {Troja}, {Uchiyama}, {Usher}, {Vandenbroucke}, {Vasileiou}, {Vianello}, {Vitale}, {Waite}, {Wang}, {Winer}, {Wood}, {Wood}, {Yang}, \& {Zimmer}}]{abdo+12}
{Abdo}, A.~A., {Ackermann}, M., {Ajello}, M., {et~al.} 2012, \apj, 758, 140, \dodoi{10.1088/0004-637X/758/2/140}

\bibitem[{{Ackermann} {et~al.}(2012){Ackermann}, {Ajello}, {Albert}, {Allafort}, {Atwood}, {Axelsson}, {Baldini}, {Ballet}, {Barbiellini}, {Bastieri}, {Bechtol}, {Bellazzini}, {Bissaldi}, {Blandford}, {Bloom}, {Bogart}, {Bonamente}, {Borgland}, {Bottacini}, {Bouvier}, {Brandt}, {Bregeon}, {Brigida}, {Bruel}, {Buehler}, {Burnett}, {Buson}, {Caliandro}, {Cameron}, {Caraveo}, {Casandjian}, {Cavazzuti}, {Cecchi}, {{\c{C}}elik}, {Charles}, {Chaves}, {Chekhtman}, {Cheung}, {Chiang}, {Ciprini}, {Claus}, {Cohen-Tanugi}, {Conrad}, {Corbet}, {Cutini}, {D'Ammando}, {Davis}, {de Angelis}, {DeKlotz}, {de Palma}, {Dermer}, {Digel}, {Silva}, {Drell}, {Drlica-Wagner}, {Dubois}, {Favuzzi}, {Fegan}, {Ferrara}, {Focke}, {Fortin}, {Fukazawa}, {Funk}, {Fusco}, {Gargano}, {Gasparrini}, {Gehrels}, {Giebels}, {Giglietto}, {Giordano}, {Giroletti}, {Glanzman}, {Godfrey}, {Grenier}, {Grove}, {Guiriec}, {Hadasch}, {Hayashida}, {Hays}, {Horan}, {Hou}, {Hughes}, {Jackson}, {Jogler}, {J{\'o}hannesson}, {Johnson}, {Johnson}, {Johnson},
  {Kamae}, {Katagiri}, {Kataoka}, {Kerr}, {Kn{\"o}dlseder}, {Kuss}, {Lande}, {Larsson}, {Latronico}, {Lavalley}, {Lemoine-Goumard}, {Longo}, {Loparco}, {Lott}, {Lovellette}, {Lubrano}, {Mazziotta}, {McConville}, {McEnery}, {Mehault}, {Michelson}, {Mitthumsiri}, {Mizuno}, {Moiseev}, {Monte}, {Monzani}, {Morselli}, {Moskalenko}, {Murgia}, {Naumann-Godo}, {Nemmen}, {Nishino}, {Norris}, {Nuss}, {Ohno}, {Ohsugi}, {Okumura}, {Omodei}, {Orienti}, {Orlando}, {Ormes}, {Paneque}, {Panetta}, {Perkins}, {Pesce-Rollins}, {Pierbattista}, {Piron}, {Pivato}, {Porter}, {Racusin}, {Rain{\`o}}, {Rando}, {Razzano}, {Razzaque}, {Reimer}, {Reimer}, {Reposeur}, {Reyes}, {Ritz}, {Rochester}, {Romoli}, {Roth}, {Sadrozinski}, {Sanchez}, {Saz Parkinson}, {Sbarra}, {Scargle}, {Sgr{\`o}}, {Siegal-Gaskins}, {Siskind}, {Spandre}, {Spinelli}, {Stephens}, {Suson}, {Tajima}, {Takahashi}, {Tanaka}, {Thayer}, {Thayer}, {Thompson}, {Tibaldo}, {Tinivella}, {Tosti}, {Troja}, {Usher}, {Vandenbroucke}, {Van Klaveren}, {Vasileiou}, {Vianello},
  {Vitale}, {Waite}, {Wallace}, {Winer}, {Wood}, {Wood}, {Wood}, {Yang}, \& {Zimmer}}]{ackermann+12}
{Ackermann}, M., {Ajello}, M., {Albert}, A., {et~al.} 2012, \apjs, 203, 4, \dodoi{10.1088/0067-0049/203/1/4}

\bibitem[{{Ackermann} {et~al.}(2016){Ackermann}, {Ajello}, {Albert}, {Atwood}, {Baldini}, {Barbiellini}, {Bastieri}, {Bellazzini}, {Bissaldi}, {Blandford}, {Bonino}, {Bottacini}, {Bregeon}, {Bruel}, {Buehler}, {Caliandro}, {Cameron}, {Caragiulo}, {Caraveo}, {Cavazzuti}, {Cecchi}, {Chekhtman}, {Chiang}, {Chiaro}, {Ciprini}, {Claus}, {Cohen-Tanugi}, {Costanza}, {Cuoco}, {Cutini}, {D'Ammando}, {de Angelis}, {de Palma}, {Desiante}, {Digel}, {Di Venere}, {Drell}, {Favuzzi}, {Fegan}, {Focke}, {Franckowiak}, {Funk}, {Fusco}, {Gargano}, {Gasparrini}, {Giglietto}, {Giordano}, {Giroletti}, {Glanzman}, {Godfrey}, {Grenier}, {Grove}, {Guiriec}, {Harding}, {Hewitt}, {Horan}, {Hou}, {Iafrate}, {J{\'o}hannesson}, {Kamae}, {Kuss}, {Larsson}, {Latronico}, {Li}, {Li}, {Longo}, {Loparco}, {Lovellette}, {Lubrano}, {Magill}, {Maldera}, {Manfreda}, {Mayer}, {Mazziotta}, {Michelson}, {Mitthumsiri}, {Mizuno}, {Monzani}, {Morselli}, {Murgia}, {Nuss}, {Omodei}, {Orlando}, {Ormes}, {Paneque}, {Perkins}, {Pesce-Rollins}, {Petrosian},
  {Piron}, {Pivato}, {Rain{\`o}}, {Rando}, {Razzano}, {Reimer}, {Reimer}, {Reposeur}, {Sgr{\`o}}, {Siskind}, {Spada}, {Spandre}, {Spinelli}, {Takahashi}, {Thayer}, {Thompson}, {Tibaldo}, {Torres}, {Tosti}, {Troja}, {Vianello}, {Winer}, {Wood}, {Yassine}, {Cerutti}, {Ferrari}, {Sala}, \& {Fermi LAT Collaboration}}]{ackermann+16}
---. 2016, \prd, 93, 082001, \dodoi{10.1103/PhysRevD.93.082001}

\bibitem[{{Agostinelli} {et~al.}(2003){Agostinelli}, {Allison}, {Amako}, {Apostolakis}, {Araujo}, {Arce}, {Asai}, {Axen}, {Banerjee}, {Barrand}, {Behner}, {Bellagamba}, {Boudreau}, {Broglia}, {Brunengo}, {Burkhardt}, {Chauvie}, {Chuma}, {Chytracek}, {Cooperman}, {Cosmo}, {Degtyarenko}, {Dell'Acqua}, {Depaola}, {Dietrich}, {Enami}, {Feliciello}, {Ferguson}, {Fesefeldt}, {Folger}, {Foppiano}, {Forti}, {Garelli}, {Giani}, {Giannitrapani}, {Gibin}, {G{\'o}mez Cadenas}, {Gonz{\'a}lez}, {Gracia Abril}, {Greeniaus}, {Greiner}, {Grichine}, {Grossheim}, {Guatelli}, {Gumplinger}, {Hamatsu}, {Hashimoto}, {Hasui}, {Heikkinen}, {Howard}, {Ivanchenko}, {Johnson}, {Jones}, {Kallenbach}, {Kanaya}, {Kawabata}, {Kawabata}, {Kawaguti}, {Kelner}, {Kent}, {Kimura}, {Kodama}, {Kokoulin}, {Kossov}, {Kurashige}, {Lamanna}, {Lamp{\'e}n}, {Lara}, {Lefebure}, {Lei}, {Liendl}, {Lockman}, {Longo}, {Magni}, {Maire}, {Medernach}, {Minamimoto}, {Mora de Freitas}, {Morita}, {Murakami}, {Nagamatu}, {Nartallo}, {Nieminen}, {Nishimura},
  {Ohtsubo}, {Okamura}, {O'Neale}, {Oohata}, {Paech}, {Perl}, {Pfeiffer}, {Pia}, {Ranjard}, {Rybin}, {Sadilov}, {Di Salvo}, {Santin}, {Sasaki}, {Savvas}, {Sawada}, {Scherer}, {Sei}, {Sirotenko}, {Smith}, {Starkov}, {Stoecker}, {Sulkimo}, {Takahata}, {Tanaka}, {Tcherniaev}, {Safai Tehrani}, {Tropeano}, {Truscott}, {Uno}, {Urban}, {Urban}, {Verderi}, {Walkden}, {Wander}, {Weber}, {Wellisch}, {Wenaus}, {Williams}, {Wright}, {Yamada}, {Yoshida}, {Zschiesche}, \& {G EANT4 Collaboration}}]{agostinelli+03}
{Agostinelli}, S., {Allison}, J., {Amako}, K., {et~al.} 2003, Nuclear Instruments and Methods in Physics Research A, 506, 250, \dodoi{10.1016/S0168-9002(03)01368-8}

\bibitem[{{Aguilar} {et~al.}(2015{\natexlab{a}}){Aguilar}, {Aisa}, {Alpat}, {Alvino}, {Ambrosi}, {Andeen}, {Arruda}, {Attig}, {Azzarello}, {Bachlechner}, {Barao}, {Barrau}, {Barrin}, {Bartoloni}, {Basara}, {Battarbee}, {Battiston}, {Bazo}, {Becker}, {Behlmann}, {Beischer}, {Berdugo}, {Bertucci}, {Bigongiari}, {Bindi}, {Bizzaglia}, {Bizzarri}, {Boella}, {de Boer}, {Bollweg}, {Bonnivard}, {Borgia}, {Borsini}, {Boschini}, {Bourquin}, {Burger}, {Cadoux}, {Cai}, {Capell}, {Caroff}, {Casaus}, {Cascioli}, {Castellini}, {Cernuda}, {Cerreta}, {Cervelli}, {Chae}, {Chang}, {Chen}, {Chen}, {Cheng}, {Chen}, {Cheng}, {Chou}, {Choumilov}, {Choutko}, {Chung}, {Clark}, {Clavero}, {Coignet}, {Consolandi}, {Contin}, {Corti}, {Gil}, {Coste}, {Creus}, {Crispoltoni}, {Cui}, {Dai}, {Delgado}, {Della Torre}, {Demirk{\"o}z}, {Derome}, {Di Falco}, {Di Masso}, {Dimiccoli}, {D{\'\i}az}, {von Doetinchem}, {Donnini}, {Du}, {Duranti}, {D'Urso}, {Eline}, {Eppling}, {Eronen}, {Fan}, {Farnesini}, {Feng}, {Fiandrini}, {Fiasson}, {Finch},
  {Fisher}, {Galaktionov}, {Gallucci}, {Garc{\'\i}a}, {Garc{\'\i}a-L{\'o}pez}, {Gargiulo}, {Gast}, {Gebauer}, {Gervasi}, {Ghelfi}, {Gillard}, {Giovacchini}, {Goglov}, {Gong}, {Goy}, {Grabski}, {Grandi}, {Graziani}, {Guandalini}, {Guerri}, {Guo}, {Haas}, {Habiby}, {Haino}, {Han}, {He}, {Heil}, {Hoffman}, {Hsieh}, {Huang}, {Huh}, {Incagli}, {Ionica}, {Jang}, {Jinchi}, {Kanishev}, {Kim}, {Kim}, {Kirn}, {Kossakowski}, {Kounina}, {Kounine}, {Koutsenko}, {Krafczyk}, {La Vacca}, {Laudi}, {Laurenti}, {Lazzizzera}, {Lebedev}, {Lee}, {Lee}, {Leluc}, {Levi}, {Li}, {Li}, {Li}, {Li}, {Li}, {Li}, {Li}, {Li}, {Li}, {Lim}, {Lin}, {Lipari}, {Lippert}, {Liu}, {Liu}, {Lolli}, {Lomtadze}, {Lu}, {Lu}, {Lu}, {Luebelsmeyer}, {Luo}, {Lv}, {Majka}, {Ma{\~n}{\'a}}, {Mar{\'\i}n}, {Martin}, {Mart{\'\i}nez}, {Masi}, {Maurin}, {Menchaca-Rocha}, {Meng}, {Mo}, {Morescalchi}, {Mott}, {M{\"u}ller}, {Ni}, {Nikonov}, {Nozzoli}, {Nunes}, {Obermeier}, {Oliva}, {Orcinha}, {Palmonari}, {Palomares}, {Paniccia}, {Papi}, {Pauluzzi}, {Pedreschi},
  {Pensotti}, {Pereira}, {Picot-Clemente}, {Pilo}, {Piluso}, {Pizzolotto}, {Plyaskin}, {Pohl}, {Poireau}, {Postaci}, {Putze}, {Quadrani}, {Qi}, {Qin}, {Qu}, {R{\"a}ih{\"a}}, {Rancoita}, {Rapin}, {Ricol}, {Rodr{\'\i}guez}, {Rosier-Lees}, {Rozhkov}, {Rozza}, {Sagdeev}, {Sandweiss}, {Saouter}, {Sbarra}, {Schael}, {Schmidt}, {von Dratzig}, {Schwering}, {Scolieri}, {Seo}, {Shan}, {Shan}, {Shi}, {Shi}, {Shi}, {Siedenburg}, {Son}, {Spada}, {Spinella}, {Sun}, {Sun}, {Tacconi}, {Tang}, {Tang}, {Tang}, {Tao}, {Tescaro}, {Ting}, {Ting}, {Tomassetti}, {Torsti}, {T{\"u}rko{\v{g}}lu}, {Urban}, {Vagelli}, {Valente}, {Vannini}, {Valtonen}, {Vaurynovich}, {Vecchi}, {Velasco}, {Vialle}, {Vitale}, {Vitillo}, {Wang}, {Wang}, {Wang}, {Wang}, {Wang}, {Wang}, {Weng}, {Whitman}, {Wienkenh{\"o}ver}, {Wu}, {Wu}, {Xia}, {Xie}, {Xie}, {Xiong}, {Xin}, {Xu}, {Xu}, {Yan}, {Yang}, {Yang}, {Ye}, {Yi}, {Yu}, {Yu}, {Zeissler}, {Zhang}, {Zhang}, {Zhang}, {Zhang}, {Zheng}, {Zhuang}, {Zhukov}, {Zichichi}, {Zimmermann}, {Zuccon}, {Zurbach}, \&
  {AMS Collaboration}}]{aguilar+15_p}
{Aguilar}, M., {Aisa}, D., {Alpat}, B., {et~al.} 2015{\natexlab{a}}, \prl, 114, 171103, \dodoi{10.1103/PhysRevLett.114.171103}

\bibitem[{{Aguilar} {et~al.}(2015{\natexlab{b}}){Aguilar}, {Aisa}, {Alpat}, {Alvino}, {Ambrosi}, {Andeen}, {Arruda}, {Attig}, {Azzarello}, {Bachlechner}, {Barao}, {Barrau}, {Barrin}, {Bartoloni}, {Basara}, {Battarbee}, {Battiston}, {Bazo}, {Becker}, {Behlmann}, {Beischer}, {Berdugo}, {Bertucci}, {Bindi}, {Bizzaglia}, {Bizzarri}, {Boella}, {de Boer}, {Bollweg}, {Bonnivard}, {Borgia}, {Borsini}, {Boschini}, {Bourquin}, {Burger}, {Cadoux}, {Cai}, {Capell}, {Caroff}, {Casaus}, {Castellini}, {Cernuda}, {Cerreta}, {Cervelli}, {Chae}, {Chang}, {Chen}, {Chen}, {Chen}, {Chen}, {Cheng}, {Chou}, {Choumilov}, {Choutko}, {Chung}, {Clark}, {Clavero}, {Coignet}, {Consolandi}, {Contin}, {Corti}, {Gil}, {Coste}, {Creus}, {Crispoltoni}, {Cui}, {Dai}, {Delgado}, {Della Torre}, {Demirk{\"o}z}, {Derome}, {Di Falco}, {Di Masso}, {Dimiccoli}, {D{\'\i}az}, {von Doetinchem}, {Donnini}, {Duranti}, {D'Urso}, {Egorov}, {Eline}, {Eppling}, {Eronen}, {Fan}, {Farnesini}, {Feng}, {Fiandrini}, {Fiasson}, {Finch}, {Fisher}, {Formato},
  {Galaktionov}, {Gallucci}, {Garc{\'\i}a}, {Garc{\'\i}a-L{\'o}pez}, {Gargiulo}, {Gast}, {Gebauer}, {Gervasi}, {Ghelfi}, {Giovacchini}, {Goglov}, {Gong}, {Goy}, {Grabski}, {Grandi}, {Graziani}, {Guandalini}, {Guerri}, {Guo}, {Haas}, {Habiby}, {Haino}, {Han}, {He}, {Heil}, {Hoffman}, {Hsieh}, {Huang}, {Huh}, {Incagli}, {Ionica}, {Jang}, {Jinchi}, {Kanishev}, {Kim}, {Kim}, {Kirn}, {Korkmaz}, {Kossakowski}, {Kounina}, {Kounine}, {Koutsenko}, {Krafczyk}, {La Vacca}, {Laudi}, {Laurenti}, {Lazzizzera}, {Lebedev}, {Lee}, {Lee}, {Leluc}, {Li}, {Li}, {Li}, {Li}, {Li}, {Li}, {Li}, {Li}, {Li}, {Li}, {Lim}, {Lin}, {Lipari}, {Lippert}, {Liu}, {Liu}, {Liu}, {Lolli}, {Lomtadze}, {Lu}, {Lu}, {Lu}, {Luebelsmeyer}, {Luo}, {Luo}, {Lv}, {Majka}, {Ma{\~n}{\'a}}, {Mar{\'\i}n}, {Martin}, {Mart{\'\i}nez}, {Masi}, {Maurin}, {Menchaca-Rocha}, {Meng}, {Mo}, {Morescalchi}, {Mott}, {M{\"u}ller}, {Nelson}, {Ni}, {Nikonov}, {Nozzoli}, {Nunes}, {Obermeier}, {Oliva}, {Orcinha}, {Palmonari}, {Palomares}, {Paniccia}, {Papi}, {Pauluzzi},
  {Pedreschi}, {Pensotti}, {Pereira}, {Picot-Clemente}, {Pilo}, {Piluso}, {Pizzolotto}, {Plyaskin}, {Pohl}, {Poireau}, {Putze}, {Quadrani}, {Qi}, {Qin}, {Qu}, {R{\"a}ih{\"a}}, {Rancoita}, {Rapin}, {Ricol}, {Rodr{\'\i}guez}, {Rosier-Lees}, {Rozhkov}, {Rozza}, {Sagdeev}, {Sandweiss}, {Saouter}, {Schael}, {Schmidt}, {von Dratzig}, {Schwering}, {Scolieri}, {Seo}, {Shan}, {Shan}, {Shi}, {Shi}, {Shi}, {Siedenburg}, {Son}, {Song}, {Spada}, {Spinella}, {Sun}, {Sun}, {Tacconi}, {Tang}, {Tang}, {Tang}, {Tao}, {Tescaro}, {Ting}, {Ting}, {Tomassetti}, {Torsti}, {T{\"u}rko{\v{g}}lu}, {Urban}, {Vagelli}, {Valente}, {Vannini}, {Valtonen}, {Vaurynovich}, {Vecchi}, {Velasco}, {Vialle}, {Vitale}, {Vitillo}, {Wang}, {Wang}, {Wang}, {Wang}, {Wang}, {Wang}, {Weng}, {Whitman}, {Wienkenh{\"o}ver}, {Willenbrock}, {Wu}, {Wu}, {Xia}, {Xie}, {Xie}, {Xiong}, {Xu}, {Xu}, {Yan}, {Yang}, {Yang}, {Yang}, {Ye}, {Yi}, {Yu}, {Yu}, {Zeissler}, {Zhang}, {Zhang}, {Zhang}, {Zhang}, {Zhang}, {Zhang}, {Zhang}, {Zheng}, {Zhuang}, {Zhukov},
  {Zichichi}, {Zimmermann}, {Zuccon}, \& {AMS Collaboration}}]{aguilar+15_a}
---. 2015{\natexlab{b}}, \prl, 115, 211101, \dodoi{10.1103/PhysRevLett.115.211101}

\bibitem[{{Allison} {et~al.}(2006){Allison}, {Amako}, {Apostolakis}, {Araujo}, {Dubois}, {Asai}, {Barrand}, {Capra}, {Chauvie}, {Chytracek}, {Cirrone}, {Cooperman}, {Cosmo}, {Cuttone}, {Daquino}, {Donszelmann}, {Dressel}, {Folger}, {Foppiano}, {Generowicz}, {Grichine}, {Guatelli}, {Gumplinger}, {Heikkinen}, {Hrivnacova}, {Howard}, {Incerti}, {Ivanchenko}, {Johnson}, {Jones}, {Koi}, {Kokoulin}, {Kossov}, {Kurashige}, {Lara}, {Larsson}, {Lei}, {Link}, {Longo}, {Maire}, {Mantero}, {Mascialino}, {McLaren}, {Lorenzo}, {Minamimoto}, {Murakami}, {Nieminen}, {Pandola}, {Parlati}, {Peralta}, {Perl}, {Pfeiffer}, {Pia}, {Ribon}, {Rodrigues}, {Russo}, {Sadilov}, {Santin}, {Sasaki}, {Smith}, {Starkov}, {Tanaka}, {Tcherniaev}, {Tome}, {Trindade}, {Truscott}, {Urban}, {Verderi}, {Walkden}, {Wellisch}, {Williams}, {Wright}, \& {Yoshida}}]{allison+06}
{Allison}, J., {Amako}, K., {Apostolakis}, J., {et~al.} 2006, IEEE Transactions on Nuclear Science, 53, 270, \dodoi{10.1109/TNS.2006.869826}

\bibitem[{{Allison} {et~al.}(2016){Allison}, {Amako}, {Apostolakis}, {Arce}, {Asai}, {Aso}, {Bagli}, {Bagulya}, {Banerjee}, {Barrand}, {Beck}, {Bogdanov}, {Brandt}, {Brown}, {Burkhardt}, {Canal}, {Cano-Ott}, {Chauvie}, {Cho}, {Cirrone}, {Cooperman}, {Cort{\'e}s-Giraldo}, {Cosmo}, {Cuttone}, {Depaola}, {Desorgher}, {Dong}, {Dotti}, {Elvira}, {Folger}, {Francis}, {Galoyan}, {Garnier}, {Gayer}, {Genser}, {Grichine}, {Guatelli}, {Gu{\`e}ye}, {Gumplinger}, {Howard}, {H{\v{r}}ivn{\'a}{\v{c}}ov{\'a}}, {Hwang}, {Incerti}, {Ivanchenko}, {Ivanchenko}, {Jones}, {Jun}, {Kaitaniemi}, {Karakatsanis}, {Karamitrosi}, {Kelsey}, {Kimura}, {Koi}, {Kurashige}, {Lechner}, {Lee}, {Longo}, {Maire}, {Mancusi}, {Mantero}, {Mendoza}, {Morgan}, {Murakami}, {Nikitina}, {Pandola}, {Paprocki}, {Perl}, {Petrovi{\'c}}, {Pia}, {Pokorski}, {Quesada}, {Raine}, {Reis}, {Ribon}, {Risti{\'c} Fira}, {Romano}, {Russo}, {Santin}, {Sasaki}, {Sawkey}, {Shin}, {Strakovsky}, {Taborda}, {Tanaka}, {Tom{\'e}}, {Toshito}, {Tran}, {Truscott}, {Urban},
  {Uzhinsky}, {Verbeke}, {Verderi}, {Wendt}, {Wenzel}, {Wright}, {Wright}, {Yamashita}, {Yarba}, \& {Yoshida}}]{allison+16}
---. 2016, Nuclear Instruments and Methods in Physics Research A, 835, 186, \dodoi{10.1016/j.nima.2016.06.125}

\bibitem[{{Anand} {et~al.}(2003){Anand}, {Taylor}, {Misra}, {Demidova}, \& {Nazarov}}]{anand+03}
{Anand}, M., {Taylor}, L.~A., {Misra}, K.~C., {Demidova}, S.~I., \& {Nazarov}, M.~A. 2003, \maps, 38, 485, \dodoi{10.1111/j.1945-5100.2003.tb00022.x}

\bibitem[{Andersson {et~al.}(1987)Andersson, Gustafson, \& Nilsson-Almqvist}]{andersson+87}
Andersson, B., Gustafson, G., \& Nilsson-Almqvist, B. 1987, Nuclear Physics B, 281, 289, \dodoi{https://doi.org/10.1016/0550-3213(87)90257-4}

\bibitem[{{Aramaki} {et~al.}(2020){Aramaki}, {Adrian}, {Karagiorgi}, \& {Odaka}}]{aramaki+20}
{Aramaki}, T., {Adrian}, P. O.~H., {Karagiorgi}, G., \& {Odaka}, H. 2020, Astroparticle Physics, 114, 107, \dodoi{10.1016/j.astropartphys.2019.07.002}

\bibitem[{Arce {et~al.}(2021)Arce, Bolst, Bordage, Brown, Cirrone, Cortés-Giraldo, Cutajar, Cuttone, Desorgher, Dondero, Dotti, Faddegon, Fedon, Guatelli, Incerti, Ivanchenko, Konstantinov, Kyriakou, Latyshev, Le, Mancini-Terracciano, Maire, Mantero, Novak, Omachi, Pandola, Perales, Perrot, Petringa, Quesada, Ramos-Méndez, Romano, Rosenfeld, Sarmiento, Sakata, Sasaki, Sechopoulos, Simpson, Toshito, \& Wright}]{arce+21}
Arce, P., Bolst, D., Bordage, M.-C., {et~al.} 2021, Medical Physics, 48, 19, \dodoi{https://doi.org/10.1002/mp.14226}

\bibitem[{Archambault {et~al.}(2003)Archambault, Beaulieu, Carrier, Castrovillari, Chauvie, Foppiano, Ghiso, Guatelli, Incerti, Lamanna, Larsson, Lopes, Peralta, Pia, Rodrigues, Tremblay, \& Trindade}]{archambault+03}
Archambault, L., Beaulieu, L., Carrier, J., {et~al.} 2003, in 2003 IEEE Nuclear Science Symposium. Conference Record (IEEE Cat. No.03CH37515), Vol.~3, 1743--1745 Vol.3, \dodoi{10.1109/NSSMIC.2003.1352215}

\bibitem[{Bennaceur \& Dobaczewski(2005)}]{bennaceur_dobaczewski05}
Bennaceur, K., \& Dobaczewski, J. 2005, Computer Physics Communications, 168, 96, \dodoi{https://doi.org/10.1016/j.cpc.2005.02.002}

\bibitem[{Bertini(1963)}]{bertini63}
Bertini, H.~W. 1963, Phys. Rev., 131, 1801, \dodoi{10.1103/PhysRev.131.1801}

\bibitem[{{Boudard} {et~al.}(2013){Boudard}, {Cugnon}, {David}, {Leray}, \& {Mancusi}}]{boudard+13}
{Boudard}, A., {Cugnon}, J., {David}, J.~C., {Leray}, S., \& {Mancusi}, D. 2013, \prc, 87, 014606, \dodoi{10.1103/PhysRevC.87.014606}

\bibitem[{Brown {et~al.}(2018)Brown, Chadwick, Capote, Kahler, Trkov, Herman, Sonzogni, Danon, Carlson, Dunn, Smith, Hale, Arbanas, Arcilla, Bates, Beck, Becker, Brown, Casperson, Conlin, Cullen, Descalle, Firestone, Gaines, Guber, Hawari, Holmes, Johnson, Kawano, Kiedrowski, Koning, Kopecky, Leal, Lestone, Lubitz, {Márquez Damián}, Mattoon, McCutchan, Mughabghab, Navratil, Neudecker, Nobre, Noguere, Paris, Pigni, Plompen, Pritychenko, Pronyaev, Roubtsov, Rochman, Romano, Schillebeeckx, Simakov, Sin, Sirakov, Sleaford, Sobes, Soukhovitskii, Stetcu, Talou, Thompson, {van der Marck}, Welser-Sherrill, Wiarda, White, Wormald, Wright, Zerkle, Žerovnik, \& Zhu}]{brown+18}
Brown, D., Chadwick, M., Capote, R., {et~al.} 2018, Nuclear Data Sheets, 148, 1, \dodoi{https://doi.org/10.1016/j.nds.2018.02.001}

\bibitem[{{Capella} {et~al.}(1994){Capella}, {Sukhatme}, {Tan}, \& {Tran Thanh Van}}]{capella+94}
{Capella}, A., {Sukhatme}, U., {Tan}, C.~I., \& {Tran Thanh Van}, J. 1994, \physrep, 236, 225, \dodoi{10.1016/0370-1573(94)90064-7}

\bibitem[{{Chatrchyan} {et~al.}(2012){Chatrchyan}, {Khachatryan}, {Sirunyan}, {Tumasyan}, {Adam}, {Aguilo}, {Bergauer}, {Dragicevic}, {Er{\"o}}, {Fabjan}, {Friedl}, {Fr{\"u}hwirth}, {Ghete}, {Hammer}, {Hoch}, {H{\"o}rmann}, {Hrubec}, {Jeitler}, {Kiesenhofer}, {Kn{\"u}nz}, {Krammer}, {Kr{\"a}tschmer}, {Liko}, {Majerotto}, {Mikulec}, {Pernicka}, {Rahbaran}, {Rohringer}, {Rohringer}, {Sch{\"o}fbeck}, {Strauss}, {Szoncs{\'o}}, {Taurok}, {Waltenberger}, {Walzel}, {Widl}, {Wulz}, {Chekhovsky}, {Emeliantchik}, {Litomin}, {Makarenko}, {Mossolov}, {Shumeiko}, {Solin}, {Stefanovitch}, {Suarez Gonzalez}, {Fedorov}, {Korzhik}, {Missevitch}, {Zuyeuski}, {Bansal}, {Bansal}, {Beaumont}, {Cornelis}, {De Wolf}, {Druzhkin}, {Janssen}, {Luyckx}, {Mucibello}, {Ochesanu}, {Roland}, {Rougny}, {Selvaggi}, {Staykova}, {Van Haevermaet}, {Van Mechelen}, {Van Remortel}, {Van Spilbeeck}, {Blekman}, {Blyweert}, {D'Hondt}, {Devroede}, {Gonzalez Suarez}, {Goorens}, {Kalogeropoulos}, {Maes}, {Olbrechts}, {Tavernier}, {Van Doninck}, {Van
  Lancker}, {Van Mulders}, {Van Onsem}, {Villella}, {Clerbaux}, {De Lentdecker}, {Dero}, {Dewulf}, {Gay}, {Hreus}, {L{\'e}onard}, {Marage}, {Mohammadi}, {Reis}, {Rugovac}, {Thomas}, {Vander Velde}, {Vanlaer}, {Wang}, {Wickens}, {Adler}, {Beernaert}, {Cimmino}, {Costantini}, {Garcia}, {Grunewald}, {Klein}, {Lellouch}, {Marinov}, {Mccartin}, {Ocampo Rios}, {Ryckbosch}, {Strobbe}, {Thyssen}, {Tytgat}, {Walsh}, {Yazgan}, {Zaganidis}, {Basegmez}, {Bruno}, {Castello}, {Ceard}, {De Favereau De Jeneret}, {Delaere}, {Demin}, {du Pree}, {Favart}, {Forthomme}, {Giammanco}, {Gr{\'e}goire}, {Hollar}, {Lemaitre}, {Liao}, {Militaru}, {Nuttens}, {Pagano}, {Pin}, {Piotrzkowski}, {Schul}, {Vizan Garcia}, {Beliy}, {Caebergs}, {Daubie}, {Hammad}, {Alves}, {Brito}, {Correa Martin}, {Martins}, {Pol}, {Souza}, {Ald{\'a} J{\'u}nior}, {Carvalho}, {Cust{\'o}dio}, {Da Costa}, {De Jesus Damiao}, {De Oliveira Martins}, {Fonseca De Souza}, {Matos Figueiredo}, {Mundim}, {Nogima}, {Oguri}, {Prado Da Silva}, {Santoro}, {Sznajder}, {Vilela
  Pereira}, {Anjos}, {Bernardes}, {Dias}, {Fernandez Perez Tomei}, {Gregores}, {Iope}, {Lagana}, {Lietti}, {Marinho}, {Mercadante}, {Novaes}, {Padula}, {Dimitrov}, {Genchev}, {Iaydjiev}, {Piperov}, {Rodozov}, {Stoykova}, {Sultanov}, {Tcholakov}, {Trayanov}, {Vankov}, {Vutova}, {Roumenin}, {Uzunova}, {Zahariev}, {Dimitrov}, {Hadjiiska}, {Kozhuharov}, {Litov}, {Pavlov}, {Petkov}, {Bian}, {Chen}, {Chen}, \& {He}}]{chatrchyan+12}
{Chatrchyan}, S., {Khachatryan}, V., {Sirunyan}, A.~M., {et~al.} 2012, Physics Letters B, 716, 30, \dodoi{10.1016/j.physletb.2012.08.021}

\bibitem[{{Cummings} \& {Stone}(2007)}]{cummings_stone_07}
{Cummings}, A.~C., \& {Stone}, E.~C. 2007, \ssr, 130, 389, \dodoi{10.1007/s11214-007-9161-y}

\bibitem[{{Cummings} {et~al.}(2002){Cummings}, {Stone}, \& {Steenberg}}]{cummings+02}
{Cummings}, A.~C., {Stone}, E.~C., \& {Steenberg}, C.~D. 2002, \apj, 578, 194, \dodoi{10.1086/342427}

\bibitem[{{Cummings} {et~al.}(2016){Cummings}, {Stone}, {Heikkila}, {Lal}, {Webber}, {J{\'o}hannesson}, {Moskalenko}, {Orlando}, \& {Porter}}]{cummings+16}
{Cummings}, A.~C., {Stone}, E.~C., {Heikkila}, B.~C., {et~al.} 2016, \apj, 831, 18, \dodoi{10.3847/0004-637X/831/1/18}

\bibitem[{{De Angelis} {et~al.}(2017){De Angelis}, {Tatischeff}, {Tavani}, {Oberlack}, {Grenier}, {Hanlon}, {Walter}, {Argan}, {von Ballmoos}, {Bulgarelli}, {Donnarumma}, {Hernanz}, {Kuvvetli}, {Pearce}, {Zdziarski}, {Aboudan}, {Ajello}, {Ambrosi}, {Bernard}, {Bernardini}, {Bonvicini}, {Brogna}, {Branchesi}, {Budtz-Jorgensen}, {Bykov}, {Campana}, {Cardillo}, {Coppi}, {De Martino}, {Diehl}, {Doro}, {Fioretti}, {Funk}, {Ghisellini}, {Grove}, {Hamadache}, {Hartmann}, {Hayashida}, {Isern}, {Kanbach}, {Kiener}, {Kn{\"o}dlseder}, {Labanti}, {Laurent}, {Limousin}, {Longo}, {Mannheim}, {Marisaldi}, {Martinez}, {Mazziotta}, {McEnery}, {Mereghetti}, {Minervini}, {Moiseev}, {Morselli}, {Nakazawa}, {Orleanski}, {Paredes}, {Patricelli}, {Peyr{\'e}}, {Piano}, {Pohl}, {Ramarijaona}, {Rando}, {Reichardt}, {Roncadelli}, {Silva}, {Tavecchio}, {Thompson}, {Turolla}, {Ulyanov}, {Vacchi}, {Wu}, \& {Zoglauer}}]{deangelis+17}
{De Angelis}, A., {Tatischeff}, V., {Tavani}, M., {et~al.} 2017, Experimental Astronomy, 44, 25, \dodoi{10.1007/s10686-017-9533-6}

\bibitem[{{De Angelis} {et~al.}(2018){De Angelis}, {Tatischeff}, {Grenier}, {McEnery}, {Mallamaci}, {Tavani}, {Oberlack}, {Hanlon}, {Walter}, {Argan}, {von Ballmoos}, {Bulgarelli}, {Bykov}, {Hernanz}, {Kanbach}, {Kuvvetli}, {Pearce}, {Zdziarski}, {Conrad}, {Ghisellini}, {Harding}, {Isern}, {Leising}, {Longo}, {Madejski}, {Martinez}, {Mazziotta}, {Paredes}, {Pohl}, {Rando}, {Razzano}, {Aboudan}, {Ackermann}, {Addazi}, {Ajello}, {Albertus}, {{\'A}lvarez}, {Ambrosi}, {Ant{\'o}n}, {Antonelli}, {Babic}, {Baibussinov}, {Balbo}, {Baldini}, {Balman}, {Bambi}, {Barres de Almeida}, {Barrio}, {Bartels}, {Bastieri}, {Bednarek}, {Bernard}, {Bernardini}, {Bernasconi}, {Bertucci}, {Biland}, {Bissaldi}, {Boettcher}, {Bonvicini}, {Bosch-Ramon}, {Bottacini}, {Bozhilov}, {Bretz}, {Branchesi}, {Brdar}, {Bringmann}, {Brogna}, {Budtz J{\o}rgensen}, {Busetto}, {Buson}, {Busso}, {Caccianiga}, {Camera}, {Campana}, {Caraveo}, {Cardillo}, {Carlson}, {Celestin}, {Cerme{\~n}o}, {Chen}, {Cheung}, {Churazov}, {Ciprini}, {Coc},
  {Colafrancesco}, {Coleiro}, {Collmar}, {Coppi}, {Curado da Silva}, {Cutini}, {D'Ammando}, {de Lotto}, {de Martino}, {De Rosa}, {Del Santo}, {Delgado}, {Diehl}, {Dietrich}, {Dolgov}, {Dom{\'\i}nguez}, {Dominis Prester}, {Donnarumma}, {Dorner}, {Doro}, {Dutra}, {Elsaesser}, {Fabrizio}, {Fern{\'a}ndez-Barral}, {Fioretti}, {Foffano}, {Formato}, {Fornengo}, {Foschini}, {Franceschini}, {Franckowiak}, {Funk}, {Fuschino}, {Gaggero}, {Galanti}, {Gargano}, {Gasparrini}, {Gehrz}, {Giammaria}, {Giglietto}, {Giommi}, {Giordano}, {Giroletti}, {Ghirlanda}, {Godinovic}, {Gouiff{\'e}s}, {Grove}, {Hamadache}, {Hartmann}, {Hayashida}, {Hryczuk}, {Jean}, {Johnson}, {Jos{\'e}}, {Kaufmann}, {Khelifi}, {Kiener}, {Kn{\"o}dlseder}, {Kole}, {Kopp}, {Kozhuharov}, {Labanti}, {Lalkovski}, {Laurent}, {Limousin}, {Linares}, {Lindfors}, {Lindner}, {Liu}, {Lombardi}, {Loparco}, {L{\'o}pez-Coto}, {L{\'o}pez Moya}, {Lott}, {Lubrano}, {Malyshev}, {Mankuzhiyil}, {Mannheim}, {March{\~a}}, {Marcian{\`o}}, {Marcote}, {Mariotti}, {Marisaldi},
  {McBreen}, {Mereghetti}, {Merle}, {Mignani}, {Minervini}, {Moiseev}, {Morselli}, {Moura}, {Nakazawa}, {Nava}, {Nieto}, {Orienti}, {Orio}, {Orlando}, {Orleanski}, {Paiano}, {Paoletti}, {Papitto}, {Pasquato}, {Patricelli}, {P{\'e}rez-Garc{\'\i}a}, {Persic}, {Piano}, {Pichel}, {Pimenta}, {Pittori}, {Porter}, {Poutanen}, {Prandini}, {Prantzos}, {Produit}, {Profumo}, {Queiroz}, {Rain{\'o}}, {Raklev}, {Regis}, {Reichardt}, {Rephaeli}, {Rico}, {Rodejohann}, {Rodriguez Fernandez}, {Roncadelli}, {Roso}, {Rovero}, {Ruffini}, {Sala}, {S{\'a}nchez-Conde}, {Santangelo}, {Saz Parkinson}, {Sbarrato}, {Shearer}, {Shellard}, {Short}, {Siegert}, {Siqueira}, {Spinelli}, {Stamerra}, {Starrfield}, {Strong}, {Str{\"u}mke}, {Tavecchio}, {Taverna}, {Terzi{\'c}}, {Thompson}, {Tibolla}, {Torres}, {Turolla}, {Ulyanov}, {Ursi}, {Vacchi}, {van den Abeele}, {Vankova-Kirilovai}, {Venter}, {Verrecchia}, {Vincent}, {Wang}, {Weniger}, {Wu}, {Zaharija{\v{s}}}, {Zampieri}, {Zane}, {Zimmer}, {Zoglauer}, \& {E-Astrogam
  Collaboration}}]{deangelis+18}
{De Angelis}, A., {Tatischeff}, V., {Grenier}, I.~A., {et~al.} 2018, Journal of High Energy Astrophysics, 19, 1, \dodoi{10.1016/j.jheap.2018.07.001}

\bibitem[{{De Angelis} {et~al.}(2021){De Angelis}, {Tatischeff}, {Argan}, {Brandt}, {Bulgarelli}, {Bykov}, {Costantini}, {Curado da Silva}, {Grenier}, {Hanlon}, {Hartmann}, {Hernanz}, {Kanbach}, {Kuvvetli}, {Laurent}, {Mazziotta}, {McEnery}, {Morselli}, {Nakazawa}, {Oberlack}, {Pearce}, {Rico}, {Tavani}, {Ballmoos}, {Walter}, {Wu}, {Zane}, {Zdziarski}, \& {Zoglauer}}]{deangelis+21}
{De Angelis}, A., {Tatischeff}, V., {Argan}, A., {et~al.} 2021, Experimental Astronomy, 51, 1225, \dodoi{10.1007/s10686-021-09706-y}

\bibitem[{{De Gaetano} {et~al.}(2023){De Gaetano}, {Di Venere}, {Gargano}, {Loparco}, {Lorusso}, {Mazziotta}, {Panzarini}, {Pillera}, \& {Serini}}]{degaetano+23}
{De Gaetano}, S., {Di Venere}, L., {Gargano}, F., {et~al.} 2023, \apj, 951, 13, \dodoi{10.3847/1538-4357/acd5ce}

\bibitem[{{Decker} {et~al.}(2005){Decker}, {Krimigis}, {Roelof}, {Hill}, {Armstrong}, {Gloeckler}, {Hamilton}, \& {Lanzerotti}}]{decker+05}
{Decker}, R.~B., {Krimigis}, S.~M., {Roelof}, E.~C., {et~al.} 2005, Science, 309, 2020, \dodoi{10.1126/science.1117569}

\bibitem[{{Desai} \& {Giacalone}(2016)}]{desai_giacalone16}
{Desai}, M., \& {Giacalone}, J. 2016, Living Reviews in Solar Physics, 13, 3, \dodoi{10.1007/s41116-016-0002-5}

\bibitem[{{Evans} {et~al.}(2006){Evans}, {Reedy}, {Starr}, {Kerry}, \& {Boynton}}]{evans+06}
{Evans}, L.~G., {Reedy}, R.~C., {Starr}, R.~D., {Kerry}, K.~E., \& {Boynton}, W.~V. 2006, Journal of Geophysical Research (Planets), 111, E03S04, \dodoi{10.1029/2005JE002657}

\bibitem[{{Fisk} {et~al.}(1974){Fisk}, {Kozlovsky}, \& {Ramaty}}]{fisk+74}
{Fisk}, L.~A., {Kozlovsky}, B., \& {Ramaty}, R. 1974, \apjl, 190, L35, \dodoi{10.1086/181498}

\bibitem[{{Fleischhack} \& {Amego X Team}(2022)}]{fleischhack+22}
{Fleischhack}, H., \& {Amego X Team}. 2022, in 37th International Cosmic Ray Conference, 649, \dodoi{10.22323/1.395.0649}

\bibitem[{Folger {et~al.}(2004)Folger, Ivanchenko, \& Wellisch}]{folger+04}
Folger, G., Ivanchenko, V.~N., \& Wellisch, J.~P. 2004, The European Physical Journal A - Hadrons and Nuclei, 21, 407, \dodoi{10.1140/epja/i2003-10219-7}

\bibitem[{{Garcia-Munoz} {et~al.}(1973){Garcia-Munoz}, {Mason}, \& {Simpson}}]{garcia-munoz+73}
{Garcia-Munoz}, M., {Mason}, G.~M., \& {Simpson}, J.~A. 1973, \apjl, 182, L81, \dodoi{10.1086/181224}

\bibitem[{{Garcia-Munoz} {et~al.}(1975){Garcia-Munoz}, {Mason}, \& {Simpson}}]{garcia-munoz+75}
---. 1975, \apj, 202, 265, \dodoi{10.1086/153973}

\bibitem[{{Giacalone} {et~al.}(2022){Giacalone}, {Fahr}, {Fichtner}, {Florinski}, {Heber}, {Hill}, {K{\'o}ta}, {Leske}, {Potgieter}, \& {Rankin}}]{giacalone+22}
{Giacalone}, J., {Fahr}, H., {Fichtner}, H., {et~al.} 2022, \ssr, 218, 22, \dodoi{10.1007/s11214-022-00890-7}

\bibitem[{{Gleeson} \& {Axford}(1968)}]{gleeson_axford68}
{Gleeson}, L.~J., \& {Axford}, W.~I. 1968, \apj, 154, 1011, \dodoi{10.1086/149822}

\bibitem[{{Gudima} {et~al.}(1983){Gudima}, {Mashnik}, \& {Toneev}}]{gudima+83}
{Gudima}, K.~K., {Mashnik}, S.~G., \& {Toneev}, V.~D. 1983, \nphysa, 401, 329, \dodoi{10.1016/0375-9474(83)90532-8}

\bibitem[{{Guthrie} {et~al.}(1968){Guthrie}, {Alsmiller}, \& {Bertini}}]{guthrie+68}
{Guthrie}, M.~P., {Alsmiller}, Jr., R.~G., \& {Bertini}, H.~W. 1968, Nuclear Instruments and Methods, 66, 29, \dodoi{10.1016/0029-554X(68)90054-2}

\bibitem[{{Hasebe} {et~al.}(2008){Hasebe}, {Shibamura}, {Miyachi}, {Takashima}, {Kobayashi}, {Okudaira}, {Yamashita}, {Kobayashi}, {Ishizaki}, {Sakurai}, {Miyajima}, {Fujii}, {Narasaki}, {Takai}, {Tsurumi}, {Kaneko}, {Nakazawa}, {Mori}, {Gasnault}, {Maurice}, {d'Uston}, {Reedy}, \& {Grande}}]{hasebe+08}
{Hasebe}, N., {Shibamura}, E., {Miyachi}, T., {et~al.} 2008, Earth, Planets and Space, 60, 299, \dodoi{10.1186/BF03352795}

\bibitem[{{Kaidalov}(1982)}]{kaidalov82}
{Kaidalov}, A.~B. 1982, Physics Letters B, 116, 459, \dodoi{10.1016/0370-2693(82)90168-X}

\bibitem[{{Kaidalov} \& {Ter-Martirosyan}(1982)}]{kaidalov_ter-martirosyan82}
{Kaidalov}, A.~B., \& {Ter-Martirosyan}, K.~A. 1982, Physics Letters B, 117, 247, \dodoi{10.1016/0370-2693(82)90556-1}

\bibitem[{{Knizhnik} {et~al.}(2011){Knizhnik}, {Swisdak}, \& {Drake}}]{knizhnik+11}
{Knizhnik}, K., {Swisdak}, M., \& {Drake}, J.~F. 2011, \apjl, 743, L35, \dodoi{10.1088/2041-8205/743/2/L35}

\bibitem[{{Kobayashi} {et~al.}(2013){Kobayashi}, {Hasebe}, {Miyachi}, {Fujii}, {Shibamura}, {Okudaira}, {Karouji}, {Hareyama}, {Takashima}, {Kobayashi}, {d'Uston}, {Maurice}, {Yamashita}, \& {Reedy}}]{kobayashi+13}
{Kobayashi}, M., {Hasebe}, N., {Miyachi}, T., {et~al.} 2013, Journal of Instrumentation, 8, P04010, \dodoi{10.1088/1748-0221/8/04/P04010}

\bibitem[{{Kobayashi} {et~al.}(2010){Kobayashi}, {Hasebe}, {Shibamura}, {Okudaira}, {Kobayashi}, {Yamashita}, {Karouji}, {Hareyama}, {Hayatsu}, {D'Uston}, {Maurice}, {Gasnault}, {Forni}, {Diez}, {Reedy}, \& {Kim}}]{kobayashi+10}
{Kobayashi}, S., {Hasebe}, N., {Shibamura}, E., {et~al.} 2010, \ssr, 154, 193, \dodoi{10.1007/s11214-010-9650-2}

\bibitem[{Koning \& Rochman(2012)}]{koning_rochman12}
Koning, A., \& Rochman, D. 2012, Nuclear Data Sheets, 113, 2841, \dodoi{https://doi.org/10.1016/j.nds.2012.11.002}

\bibitem[{Koning {et~al.}(2019)Koning, Rochman, Sublet, Dzysiuk, Fleming, \& {van der Marck}}]{koning+19}
Koning, A., Rochman, D., Sublet, J.-C., {et~al.} 2019, Nuclear Data Sheets, 155, 1, \dodoi{https://doi.org/10.1016/j.nds.2019.01.002}

\bibitem[{{Krimigis} {et~al.}(2003){Krimigis}, {Decker}, {Hill}, {Armstrong}, {Gloeckler}, {Hamilton}, {Lanzerotti}, \& {Roelof}}]{krimigis+03}
{Krimigis}, S.~M., {Decker}, R.~B., {Hill}, M.~E., {et~al.} 2003, \nat, 426, 45, \dodoi{10.1038/nature02068}

\bibitem[{{Lawrence} {et~al.}(1998){Lawrence}, {Feldman}, {Barraclough}, {Binder}, {Elphic}, {Maurice}, \& {Thomsen}}]{lawrence+98}
{Lawrence}, D.~J., {Feldman}, W.~C., {Barraclough}, B.~L., {et~al.} 1998, Science, 281, 1484, \dodoi{10.1126/science.281.5382.1484}

\bibitem[{{Lund}(1984)}]{lund84}
{Lund}, N. 1984, Advances in Space Research, 4, 5, \dodoi{10.1016/0273-1177(84)90287-4}

\bibitem[{{Ma} {et~al.}(2008){Ma}, {Chang}, {Zhang}, {Cai}, {Gong}, {Tang}, {Zhang}, {Wang}, {Yu}, {Mao}, {Su}, {Fang}, {Xu}, {Hu}, {Gu}, {Zhou}, {Xu}, \& {Liu}}]{ma+08}
{Ma}, T., {Chang}, J., {Zhang}, N., {et~al.} 2008, Advances in Space Research, 42, 347, \dodoi{10.1016/j.asr.2007.04.042}

\bibitem[{{Ma} {et~al.}(2013){Ma}, {Chang}, {Zhang}, {Jian}, {Cai}, {Gong}, {Tang}, {Zhang}, {Wang}, {Yu}, {Mao}, {Hu}, {Xu}, \& {Zhu}}]{ma+13}
---. 2013, Nuclear Instruments and Methods in Physics Research A, 726, 113, \dodoi{10.1016/j.nima.2013.05.162}

\bibitem[{{Mancusi} {et~al.}(2014){Mancusi}, {Boudard}, {Cugnon}, {David}, {Kaitaniemi}, \& {Leray}}]{davide+14}
{Mancusi}, D., {Boudard}, A., {Cugnon}, J., {et~al.} 2014, \prc, 90, 054602, \dodoi{10.1103/PhysRevC.90.054602}

\bibitem[{{McDonald} {et~al.}(2003){McDonald}, {Stone}, {Cummings}, {Heikkila}, {Lal}, \& {Webber}}]{mcdonald+03}
{McDonald}, F.~B., {Stone}, E.~C., {Cummings}, A.~C., {et~al.} 2003, \nat, 426, 48, \dodoi{10.1038/nature02066}

\bibitem[{{Mendoza} {et~al.}(2014){Mendoza}, {Cano-Ott}, {Koi}, \& {Guerrero}}]{mendoza+14}
{Mendoza}, E., {Cano-Ott}, D., {Koi}, T., \& {Guerrero}, C. 2014, IEEE Transactions on Nuclear Science, 61, 2357, \dodoi{10.1109/TNS.2014.2335538}

\bibitem[{{Metzger} {et~al.}(1973){Metzger}, {Trombka}, {Peterson}, {Reedy}, \& {Arnold}}]{metzger+73}
{Metzger}, A.~E., {Trombka}, J.~I., {Peterson}, L.~E., {Reedy}, R.~C., \& {Arnold}, J.~R. 1973, Science, 179, 800, \dodoi{10.1126/science.179.4075.800}

\bibitem[{{Morris}(1984)}]{morris84}
{Morris}, D.~J. 1984, \jgr, 89, 10685, \dodoi{10.1029/JA089iA12p10685}

\bibitem[{{Moskalenko} \& {Porter}(2007)}]{moskalenko_porter07}
{Moskalenko}, I.~V., \& {Porter}, T.~A. 2007, \apj, 670, 1467, \dodoi{10.1086/522828}

\bibitem[{{Moskalenko} {et~al.}(2008){Moskalenko}, {Porter}, {Digel}, {Michelson}, \& {Ormes}}]{moskalenko+08}
{Moskalenko}, I.~V., {Porter}, T.~A., {Digel}, S.~W., {Michelson}, P.~F., \& {Ormes}, J.~F. 2008, \apj, 681, 1708, \dodoi{10.1086/588425}

\bibitem[{{Naito} {et~al.}(2018){Naito}, {Hasebe}, {Nagaoka}, {Shibamura}, {Ohtake}, {Kim}, {W{\"o}hler}, \& {Berezhnoy}}]{naito+18}
{Naito}, M., {Hasebe}, N., {Nagaoka}, H., {et~al.} 2018, \icarus, 310, 21, \dodoi{10.1016/j.icarus.2017.12.005}

\bibitem[{Nakayama {et~al.}(2021)Nakayama, Iwamoto, Watanabe, \& Ogata}]{nakayama+21}
Nakayama, S., Iwamoto, O., Watanabe, Y., \& Ogata, K. 2021, Journal of Nuclear Science and Technology, 58, 805–821, \dodoi{10.1080/00223131.2020.1870010}

\bibitem[{{Nilsson-Almqvist} \& {Stenlund}(1987)}]{nilsson-almqvist_stenlund87}
{Nilsson-Almqvist}, B., \& {Stenlund}, E. 1987, Computer Physics Communications, 43, 387, \dodoi{10.1016/0010-4655(87)90056-7}

\bibitem[{{Odaka} {et~al.}(2018){Odaka}, {Asai}, {Hagino}, {Koi}, {Madejski}, {Mizuno}, {Ohno}, {Saito}, {Sato}, {Wright}, {Enoto}, {Fukazawa}, {Hayashi}, {Kataoka}, {Katsuta}, {Kawaharada}, {Kobayashi}, {Kokubun}, {Laurent}, {Lebrun}, {Limousin}, {Maier}, {Makishima}, {Mimura}, {Miyake}, {Mori}, {Murakami}, {Nakamori}, {Nakano}, {Nakazawa}, {Noda}, {Ohta}, {Ozaki}, {Sato}, {Sato}, {Tajima}, {Takahashi}, {Takahashi}, {Takeda}, {Tanaka}, {Tanaka}, {Terada}, {Uchiyama}, {Uchiyama}, {Watanabe}, {Yamaoka}, {Yasuda}, {Yatsu}, {Yuasa}, \& {Zoglauer}}]{odaka+18}
{Odaka}, H., {Asai}, M., {Hagino}, K., {et~al.} 2018, Nuclear Instruments and Methods in Physics Research A, 891, 92, \dodoi{10.1016/j.nima.2018.02.071}

\bibitem[{{Orlando}(2018)}]{orlando18}
{Orlando}, E. 2018, \mnras, 475, 2724, \dodoi{10.1093/mnras/stx3280}

\bibitem[{{Orlando} {et~al.}(2022){Orlando}, {Bottacini}, {Moiseev}, {Bodaghee}, {Collmar}, {Ensslin}, {Moskalenko}, {Negro}, {Profumo}, {Digel}, {Thompson}, {Baring}, {Bolotnikov}, {Cannady}, {Carini}, {Eberle}, {Grenier}, {Harding}, {Hartmann}, {Herrmann}, {Kerr}, {Krivonos}, {Laurent}, {Longo}, {Morselli}, {Philips}, {Sasaki}, {Shawhan}, {Shy}, {Skinner}, {Smith}, {Stecker}, {Strong}, {Sturner}, {Tomsick}, {Wadiasingh}, {Woolf}, {Yates}, {Ziock}, \& {Zoglauer}}]{orlando+22}
{Orlando}, E., {Bottacini}, E., {Moiseev}, A.~A., {et~al.} 2022, \jcap, 2022, 036, \dodoi{10.1088/1475-7516/2022/07/036}

\bibitem[{{Porter} {et~al.}(2017){Porter}, {J{\'o}hannesson}, \& {Moskalenko}}]{porter+17}
{Porter}, T.~A., {J{\'o}hannesson}, G., \& {Moskalenko}, I.~V. 2017, \apj, 846, 67, \dodoi{10.3847/1538-4357/aa844d}

\bibitem[{{Porter} {et~al.}(2022){Porter}, {J{\'o}hannesson}, \& {Moskalenko}}]{porter+22}
---. 2022, \apjs, 262, 30, \dodoi{10.3847/1538-4365/ac80f6}

\bibitem[{{Prettyman} {et~al.}(2006){Prettyman}, {Hagerty}, {Elphic}, {Feldman}, {Lawrence}, {McKinney}, \& {Vaniman}}]{prettyman+06}
{Prettyman}, T.~H., {Hagerty}, J.~J., {Elphic}, R.~C., {et~al.} 2006, Journal of Geophysical Research (Planets), 111, E12007, \dodoi{10.1029/2005JE002656}

\bibitem[{{Prinz} {et~al.}(1973){Prinz}, {Dowty}, {Keil}, \& {Bunch}}]{prinz+73}
{Prinz}, M., {Dowty}, E., {Keil}, K., \& {Bunch}, T.~E. 1973, \gca, 37, 979, \dodoi{10.1016/0016-7037(73)90195-6}

\bibitem[{{Ramaty} {et~al.}(1979){Ramaty}, {Kozlovsky}, \& {Lingenfelter}}]{ramaty+79}
{Ramaty}, R., {Kozlovsky}, B., \& {Lingenfelter}, R.~E. 1979, \apjs, 40, 487, \dodoi{10.1086/190596}

\bibitem[{{Reames}(1999)}]{reames99}
{Reames}, D.~V. 1999, \ssr, 90, 413, \dodoi{10.1023/A:1005105831781}

\bibitem[{{Reames} \& {Ng}(2010)}]{reames_ng_10}
{Reames}, D.~V., \& {Ng}, C.~K. 2010, \apj, 723, 1286, \dodoi{10.1088/0004-637X/723/2/1286}

\bibitem[{{Reid}(1974)}]{reid74}
{Reid}, A.~M. 1974, Moon, 9, 141, \dodoi{10.1007/BF00565400}

\bibitem[{Rodr\'{\i}guez-S\'anchez {et~al.}(2017)Rodr\'{\i}guez-S\'anchez, David, Mancusi, Boudard, Cugnon, \& Leray}]{rodriguez+17}
Rodr\'{\i}guez-S\'anchez, J.~L., David, J.-C., Mancusi, D., {et~al.} 2017, Phys. Rev. C, 96, 054602, \dodoi{10.1103/PhysRevC.96.054602}

\bibitem[{{Sch{\"o}nfelder} {et~al.}(1984){Sch{\"o}nfelder}, {Diehl}, {Lichti}, {Steinle}, {Swanenburg}, {Deerenberg}, {Aarts}, {Lockwood}, {Webber}, {Macri}, {Ryan}, {Simpson}, {Taylor}, {Bennett}, \& {Snelling}}]{schonfelder+84}
{Sch{\"o}nfelder}, V., {Diehl}, R., {Lichti}, G.~G., {et~al.} 1984, IEEE Transactions on Nuclear Science, 1, 766, \dodoi{10.1109/TNS.1984.4333363}

\bibitem[{{Shutt} {et~al.}(2025){Shutt}, {Trbalic}, {Charles}, {Di Lalla}, {Hitchcock}, {Jett}, {Linehan}, {Luitz}, {Madejski}, {Pe{\~n}a-Perez}, \& {Tsai}}]{shutt+25}
{Shutt}, T., {Trbalic}, B., {Charles}, E., {et~al.} 2025, arXiv e-prints, arXiv:2502.14841, \dodoi{10.48550/arXiv.2502.14841}

\bibitem[{{Siegert}(2024)}]{siegert+24}
{Siegert}, T. 2024, \mnras, 533, 165, \dodoi{10.1093/mnras/stae1742}

\bibitem[{{Stone} {et~al.}(2005){Stone}, {Cummings}, {McDonald}, {Heikkila}, {Lal}, \& {Webber}}]{stone+05}
{Stone}, E.~C., {Cummings}, A.~C., {McDonald}, F.~B., {et~al.} 2005, Science, 309, 2017, \dodoi{10.1126/science.1117684}

\bibitem[{{Stone} {et~al.}(2008){Stone}, {Cummings}, {McDonald}, {Heikkila}, {Lal}, \& {Webber}}]{stone+08}
---. 2008, \nat, 454, 71, \dodoi{10.1038/nature07022}

\bibitem[{{Stone} {et~al.}(2013){Stone}, {Cummings}, {McDonald}, {Heikkila}, {Lal}, \& {Webber}}]{stone+13}
---. 2013, Science, 341, 150, \dodoi{10.1126/science.1236408}

\bibitem[{{Strong} \& {Moskalenko}(1998)}]{strong_moskalenko98}
{Strong}, A.~W., \& {Moskalenko}, I.~V. 1998, \apj, 509, 212, \dodoi{10.1086/306470}

\bibitem[{{Strong} {et~al.}(2007){Strong}, {Moskalenko}, \& {Ptuskin}}]{strong+07}
{Strong}, A.~W., {Moskalenko}, I.~V., \& {Ptuskin}, V.~S. 2007, Annual Review of Nuclear and Particle Science, 57, 285, \dodoi{10.1146/annurev.nucl.57.090506.123011}

\bibitem[{{Takahashi} {et~al.}(2013){Takahashi}, {Uchiyama}, \& {Stawarz}}]{takahashi+13}
{Takahashi}, T., {Uchiyama}, Y., \& {Stawarz}, {\L}. 2013, Astroparticle Physics, 43, 142, \dodoi{10.1016/j.astropartphys.2012.05.010}

\bibitem[{{Taylor}(1975)}]{taylor75}
{Taylor}, S.~R. 1975, {Lunar science: a post-Apollo view; scientific results and insights from the lunar samples.}

\bibitem[{{Thompson} {et~al.}(1997){Thompson}, {Bertsch}, {Morris}, \& {Mukherjee}}]{thompson+97}
{Thompson}, D.~J., {Bertsch}, D.~L., {Morris}, D.~J., \& {Mukherjee}, R. 1997, \jgr, 102, 14735, \dodoi{10.1029/97JA01045}

\bibitem[{{Tomsick} {et~al.}(2024){Tomsick}, {Boggs}, {Zoglauer}, {Hartmann}, {Ajello}, {Burns}, {Fryer}, {Karwin}, {Kierans}, {Lowell}, {Malzac}, {Roberts}, {Saint-Hilaire}, {Shih}, {Siegert}, {Sleator}, {Takahashi}, {Tavecchio}, {Wulf}, {Beechert}, {Gulick}, {Joens}, {Lazar}, {Neights}, {Martinez Oliveros}, {Matsumoto}, {Melia}, {Yoneda}, {Amman}, {Bal}, {von Ballmoos}, {Bates}, {B{\"o}ttcher}, {Bulgarelli}, {Cavazzuti}, {Chang}, {Chen}, {Chu}, {Ciabattoni}, {Costamante}, {Dreyer}, {Fioretti}, {Fenu}, {Gallego}, {Ghirlanda}, {Grove}, {Huang}, {Jean}, {Khatiya}, {Kn{\"o}dlseder}, {Kraus}, {Leising}, {Lewis}, {Lommler}, {Marcotulli}, {Martinez Castellanos}, {Mittal}, {Negro}, {Al Nussirat}, {Nakazawa}, {Oberlack}, {Palmore}, {Panebianco}, {Parmiggiani}, {Pike}, {Rogers}, {Schutte}, {Sheng}, {Smale}, {Smith}, {Trigg}, {Venters}, {Watanabe}, \& {Zhang}}]{tomsick+23}
{Tomsick}, J., {Boggs}, S., {Zoglauer}, A., {et~al.} 2024, in 38th International Cosmic Ray Conference, 745, \dodoi{10.48550/arXiv.2308.12362}

\bibitem[{{Tsuji} {et~al.}(2023){Tsuji}, {Inoue}, {Yoneda}, {Mukherjee}, \& {Odaka}}]{tsuji+23}
{Tsuji}, N., {Inoue}, Y., {Yoneda}, H., {Mukherjee}, R., \& {Odaka}, H. 2023, \apj, 943, 48, \dodoi{10.3847/1538-4357/acab69}

\bibitem[{{Vedrenne} {et~al.}(2003){Vedrenne}, {Roques}, {Sch{\"o}nfelder}, {Mandrou}, {Lichti}, {von Kienlin}, {Cordier}, {Schanne}, {Kn{\"o}dlseder}, {Skinner}, {Jean}, {Sanchez}, {Caraveo}, {Teegarden}, {von Ballmoos}, {Bouchet}, {Paul}, {Matteson}, {Boggs}, {Wunderer}, {Leleux}, {Weidenspointner}, {Durouchoux}, {Diehl}, {Strong}, {Cass{\'e}}, {Clair}, \& {Andr{\'e}}}]{vedrenne+03}
{Vedrenne}, G., {Roques}, J.~P., {Sch{\"o}nfelder}, V., {et~al.} 2003, \aap, 411, L63, \dodoi{10.1051/0004-6361:20031482}

\bibitem[{{Vladimirov} {et~al.}(2011){Vladimirov}, {Digel}, {J{\'o}hannesson}, {Michelson}, {Moskalenko}, {Nolan}, {Orlando}, {Porter}, \& {Strong}}]{vladimirov+11}
{Vladimirov}, A.~E., {Digel}, S.~W., {J{\'o}hannesson}, G., {et~al.} 2011, Computer Physics Communications, 182, 1156, \dodoi{10.1016/j.cpc.2011.01.017}

\bibitem[{{Wright} \& {Kelsey}(2015)}]{wright_kelsey15}
{Wright}, D.~H., \& {Kelsey}, M.~H. 2015, Nuclear Instruments and Methods in Physics Research A, 804, 175, \dodoi{10.1016/j.nima.2015.09.058}

\bibitem[{{Yamashita} {et~al.}(2015){Yamashita}, {Prettyman}, \& {Reedy}}]{yamashita+15}
{Yamashita}, N., {Prettyman}, T.~H., \& {Reedy}, R.~C. 2015, in 46th Annual Lunar and Planetary Science Conference, Lunar and Planetary Science Conference, 2223

\bibitem[{{Yamashita} {et~al.}(2010){Yamashita}, {Hasebe}, {Reedy}, {Kobayashi}, {Karouji}, {Hareyama}, {Shibamura}, {Kobayashi}, {Okudaira}, {d'Uston}, {Gasnault}, {Forni}, \& {Kim}}]{yamashita+10}
{Yamashita}, N., {Hasebe}, N., {Reedy}, R.~C., {et~al.} 2010, \grl, 37, L10201, \dodoi{10.1029/2010GL043061}

\bibitem[{{Yamashita} {et~al.}(2012){Yamashita}, {Gasnault}, {Forni}, {d'Uston}, {Reedy}, {Karouji}, {Kobayashi}, {Hareyama}, {Nagaoka}, {Hasebe}, \& {Kim}}]{yamashita+12}
{Yamashita}, N., {Gasnault}, O., {Forni}, O., {et~al.} 2012, Earth and Planetary Science Letters, 353, 93, \dodoi{10.1016/j.epsl.2012.08.010}

\bibitem[{{Zoglauer} \& {Kanbach}(2003)}]{zoglauer_kanbach03}
{Zoglauer}, A., \& {Kanbach}, G. 2003, in Society of Photo-Optical Instrumentation Engineers (SPIE) Conference Series, Vol. 4851, X-Ray and Gamma-Ray Telescopes and Instruments for Astronomy., ed. J.~E. {Truemper} \& H.~D. {Tananbaum}, 1302--1309, \dodoi{10.1117/12.461177}

\end{thebibliography}
\bibliographystyle{aasjournal}



\end{document}